\documentclass[12pt]{iopart}

\usepackage{amssymb,amsbsy}
\expandafter\let\csname equation*\endcsname\relax
\expandafter\let\csname endequation*\endcsname\relax

\usepackage{amsmath}
\usepackage{latexsym}
 \usepackage{setspace}
\usepackage{graphicx}
\usepackage{cite}
\usepackage{hyperref}

\usepackage{tikz}
\usetikzlibrary{calc,decorations.pathreplacing}

\def\mz{\mathfrak{z}}

\def\cH{{\cal H}}
\def\cM{{\cal M}}
\def\cZ{{\cal Z}}

\def\cF{{\cal F}}
\def\cC{{\cal C}}
\def\cN{{\cal N}}
\def\cO{{\cal O}}
\def\cT{{\cal T}}
\def\N{{\mathbb N}}
\def\cE{{\cal E}}
\def\ZZ{{\mathbb Z}}
\def\Z{{\mathbb Z}}
\def\R{{\mathbb R}}
\def\cD{{\cal D}}

\def\half{{\scriptstyle \frac 12}}

\def\threeh{{\scriptstyle \frac 32}}
\def\fiveh{{\scriptstyle \frac 52}}

\newcommand{\bea}{\begin{eqnarray}}
\newcommand{\eea}{\end{eqnarray}}

\newenvironment{psmallmatrix}
  {\left(\begin{smallmatrix}}
  {\end{smallmatrix}\right)}

\def\flux{{\tilde{N}}_{G_N}}
\def\fluxSU{{\tilde{N}}_{SU(N)}}
\def\fluxSO{{\tilde{N}}_{SO(n)}}
\def\fluxUSp{{\tilde{N}}_{USp(n)}}

\usepackage{etoolbox}

\def\otau{{\pmb{\tau}}}

\makeatletter
\def\@mkboth#1#2{}
\newlength\appendixwidth
\preto\appendix{\addtocontents{toc}{\protect\patchl@section}}
\newcommand{\patchl@section}{%
  \settowidth{\appendixwidth}{\textbf{Appendix }}%
  \addtolength{\appendixwidth}{1.5em}%
  \patchcmd{\l@section}{1.5em}{\appendixwidth}{}{\ddt}%
}
\makeatother

\begin{document}
\bibliographystyle{iopart-num}

%
\newcommand{\eprint}[2][]{\href{https://arxiv.org/abs/#2}{#2}}
%

\begin{flushright}
	SAGEX-22-11  \\
	QMUL-PH-22-11
\end{flushright}

\title[Superstring amplitudes \& Integrated correlators]{The SAGEX Review on Scattering Amplitudes, Chapter 10: Selected topics on modular covariance of type IIB string amplitudes and their $\mathcal{N}=4$ supersymmetric Yang--Mills duals}

\author{Daniele Dorigoni$^{(a)}$, Michael B. Green$^{(b), (c)}$  and Congkao Wen$^{(c)}$} 
\vskip 0.1in

\address{{\small ($a$) Centre for Particle Theory \& Department of Mathematical Sciences, 
}\\
\small{Durham University, Lower Mountjoy, Stockton Road, Durham DH1 3LE, UK} 

\vskip 0.1in

{ \small ($b$) Department of Applied Mathematics and Theoretical Physics }\\
{\small  Wilberforce Road, Cambridge CB3 0WA, UK} 

\vskip 0.1in

{\small  ($c$) Centre for Theoretical Physics, Department of Physics and Astronomy,  }\\ 
{\small Queen Mary University of London,  London, E1 4NS, UK}
}
\ead{daniele.dorigoni@durham.ac.uk, M.B.Green@damtp.cam.ac.uk, c.wen@qmul.ac.uk}
\vspace{10pt}
\begin{indented}
\item[]September 2022
\end{indented}

\begin{abstract}
This article reviews some results of the  SAGEX programme that have developed in  the understanding of the interplay of supersymmetry and modular covariance of scattering amplitudes in type IIB superstring theory and its holographic image in $\mathcal{N}=4$ supersymmetric Yang--Mills theory (SYM).  The first section includes the determination of exact expressions for BPS interactions in the low-energy expansion of type IIB superstring amplitudes.  
The second section concerns properties of a certain class of integrated correlators in $\mathcal{N}=4$ SYM with arbitrary classical gauge group  that are exactly determined by supersymmetric localisation.   Not only do these reproduce known features of perturbative and non-perturbative $\mathcal{N}=4$ SYM for any classical gauge group, but they have large-$N$ expansions that are in accord with expectations based on the holographic correspondence with superstring theory.
The final section focusses on modular graph functions.  These are modular functions that are closely associated with coefficients in the low-energy expansion of superstring perturbation theory and have recently received quite a lot of interest in both the physics and mathematics literature.
\end{abstract}

%
%
%
%
%

\newpage
\tableofcontents
\newpage

\section{Introduction}
\label{sec:intro}

This article reviews recent developments  concerning properties of superstring scattering amplitudes and their relation to correlation functions of gauge-invariant composite operators in $\cN=4$ supersymmetric Yang--Mills theory (SYM). These are areas in which there has been a large amount of recent work, but we will concentrate on rather restricted features that are close to our own research interests and hopefully illustrate more general principles.  

Section~\ref{sec:susymod} is concerned with aspects of the low-energy expansion of type IIB superstring amplitudes 
 which are highly constrained by maximal supersymmetry and $SL(2,\Z)$ S-duality.
 Successive terms in this expansion are higher-dimension interactions with coefficients  that are modular forms that transform with specific holomorphic and anti-holomorphic weights under $SL(2,\Z)$ transformation of the complex scalar field $\tau=\tau_1+i \tau_2$ that parameterises the coset space $SL(2,\R)/U(1)$.
 
  The interactions that contribute to the four-graviton amplitude up to mass dimension $14$ are fractional BPS terms and are proportional to $d^{2n}R^4$, where $n\leq 3$ and  $R^4$ is a particular contraction of four Riemann tensors that is fixed by supersymmetry.  These terms have coefficients that are modular functions of $\tau$ that are fully determined by supersymmetry and S-duality supplemented by a boundary condition in the large-$\tau_2$ limit, i.e. in the weak string coupling limit.  The  coefficients of  the $R^4$ and $d^4R^4$ interactions will be seen to satisfy Laplace eigenvalue equations in the upper-half $\tau$ plane.   The solutions of these equations are known to be non-holomorphic Eisenstein series.  
  
   The coefficient of the $d^6R^4$ interaction satisfies an inhomogeneous Laplace eigenvalue equation with a source term that is quadratic in non-holomorphic Eisenstein series.  This has a solution that is a `generalised' non-holomorphic Eisenstein series.   We will see that there are many other BPS  higher-derivative  interactions that have mass dimensions $\le 14$ that are related to the four-point amplitude.  In general $n$-point amplitudes with $n>4$ may violate the conservation of the $U(1)$ R-symmetry of type IIB supergravity, due to stringy corrections.  Their coefficients transform as modular forms with  
 holomorphic and anti-holomorphic weights $(w,-w)$. Here we will consider the special amplitudes that violate  $U(1)$ R-symmetry maximally, which are known as maximal $U(1)$-violating (MUV) amplitudes and for which $w=n-4$.
 The expressions for these modular forms are determined by soft-dilaton relations combined with supersymmetry.  A summary of some mathematical properties  of modular forms, non-holomorphic Eisenstein series and generalised Eisenstein series is given in  \ref{sec:math}.

 In section~\ref{sec:correlator}  we will consider exact properties of  integrated correlators of BPS operators in the stress tensor supermultiplet of  $\cN=4$ supersymmetric Yang--Mills (SYM) theory with any classical gauge group, $G_N=SU(N)$, $SO(2N)$, $SO(2N+1)$, $USp(2N)$. These  integrated correlators are determined by the partition function of the  $\cN=2^*$ SYM theory, which can be viewed as a mass deformation of the $\cN=4$ theory. Most of our discussion will be concerned with  an integrated correlator that  is proportional to  $\Delta_\tau \partial_m^2\log Z_{G_N}(m,\tau,\bar\tau)|_{m=0}$, where the  $\cN=2^*$ partition function, $ Z(m,\tau,\bar\tau)$, is determined by supersymmetric localisation  on $S^4$, the parameter $m$ is the  hypermultiplet mass and $\Delta_\tau= 4\tau_2^2 \partial_\tau\partial_{\bar\tau}$ is the hyperbolic laplacian on the upper-half $\tau$ plane.  
 
 We will see that this correlator can be expressed as a two-dimensional lattice sum for any of the gauge groups.  This is a well-defined expression valid for all values of $N$ and $\tau$. It can also be written as a formal infinite sum of non-holomorphic Eisenstein series  of integer index with rational coefficients.  The integrated correlator with $G_N$ gauge group satisfies a rather remarkable `Laplace-difference' equation that has an iterative solution relating it to the  integrated  correlator with $SU(2)$ gauge group. The perturbative and non-perturbative instanton contributions to this integrated correlator are easy to extract for any finite value of $N$ and display a number of intriguing features. 
 
 For example,  the planar contributions  to the perturbative expansion  in powers of a suitable 't Hooft coupling $a_{G_N}$  are the same for all gauge groups and non-planar terms first enter at order $a^4_{G_N}$.  The large-$N$ expansion shows similarly interesting regularities.  Furthermore it has a structure that makes manifest the holographic relationship to the low-energy expansion of type IIB superstring theory in $AdS_5\times S^5$ in the $SU(N)$ case and  $AdS_5\times S^5/ \mathbb{Z}_2$ for other classical gauge groups. 
  Some of the large-$N$ properties of a second integrated correlator that is proportional to    $\partial_m^4\log Z_{G_N}(m,\tau,\bar\tau)|_{m=0}$ are determined in section~\ref{sec:correlator}, where the generalisation to $n$-point MUV integrated correlators is also presented. The large-$N$ expansion of these integrated correlators will be used to determine the low-energy expansion of superstring amplitudes in $AdS_5\times S^5$. In the flat-space limit, these reproduce the exact results obtained in section~\ref{sec:susymod} using different methods.
 
 Section~\ref{sec:otherasp} focusses on  properties of modular graph functions.  These modular functions are closely associated with the low-energy expansions of the perturbative contributions to type IIB superstring amplitudes.  The contribution to the amplitude at order $g_s^{2g-2}$ (where $g_s$ is the string coupling constant) is defined by a functional integral over genus-$g$ world-sheets.   
  The low-energy expansion of the tree amplitudes (the $g=0$ case) has been extensively studied and generates infinite series of powers of Mandelstam invariants with coefficients that are rational multiples of single-valued multiple zeta values.  These are special values of single-valued multiple polylogarithms, which have close connections with mathematical aspects of Feynman diagrams.  Much less is known about the general structure of $n$-point amplitudes at genus $g\ge 1$. 
 
 In general, the integration over the positions of the punctures, i.e. the vertex operators insertion points, cannot be done exactly but can be performed order by order in the low-energy, $\alpha'\to0$, expansion. The result of integrating over the positions of the punctures are functions of the world-sheet moduli.  
 In such cases the low-energy expansion is a series of terms with coefficients that are integrals of genus-$g$ `modular graph functions'. In section~\ref{sec:otherasp} we will review the structure and properties of genus-one, $g=1$, modular graph functions.  These are functions of a single complex modulus which is the complex structure of the toroidal genus-one world-sheet.
 
  These functions are elliptic generalisations of single-valued multiple zeta values that may be described in terms of Feynman diagrams for free scalar fields propagating  on a two dimensional torus.  Consequently the loop momenta are integers and the diagrams are expressed as multiple lattice sums.  The modular graph functions that are generated by the low-energy expansion of the four-point genus-one amplitude form a special subset of general genus-one modular graph functions. We will further see that a systematic analysis of their properties requires the consideration of modular graph forms that transform with non-trivial holomorphic and anti-holomorphic modular weights.   
The genus-one modular graph functions  described by two-loop Feynman diagrams  are closely related to generalised Eisenstein series defined in \ref{sec:math}.
In the final part of  section~\ref{sec:otherasp}  we will briefly describe some features of genus-two modular graph functions, which are functions of the three complex moduli of genus-two Riemann surfaces.

We end with some brief comments in section~\ref{sec:conclude}.

Due to space constraints, the description of these topics is necessarily superficial.  However, our discussion points  towards the relevant references for  those who are keen on understanding these results at a deeper level.

 \section{Supersymmetry and modular constraints on low-energy expansion}
\label{sec:susymod}

\subsection{Low-energy expansion of type IIB superstring theory}
\label{sec:4pts}

In this section we will review some exact results concerning the low-energy expansion of massless scattering amplitudes in type IIB superstring theory. The results may be conveniently expressed in terms of an effective Lagrangian that has the form (in the string frame)
\begin{eqnarray}
 \label{eq:LLeff}
\mathcal{L}_{\rm eff} &=& (\alpha^{\prime})^{-4} g_s^{-2}\,R + E(\threeh ; \tau,\bar{\tau}) (\alpha^{\prime})^{-1} g_s^{-\frac{1}{2}}R^4 + E(\fiveh; \tau,\bar{\tau}) \, \alpha^{\prime} g_s^{\frac{1}{2}}  d^4R^4 \cr
&+& \mathcal{E}(\threeh, \threeh; 3; \tau,\bar{\tau}) (\alpha^{\prime})^2 g_s\, d^6R^4  + \ldots\, . 
\end{eqnarray}
In this expression $\alpha' = \ell_s^2$ is the square of the string length scale.
The leading term proportional to  $R$ is the Einstein-Hilbert term (where $R$ is the Ricci scalar), which has mass dimension $2$. Together with its supersymmetric completion that involves more bosonic and fermionic fields, it describes type IIB supergravity in ten dimensions. The scalar field $\tau$ parameterises the coset space  $SL(2,\R)/U(1)$ in the classical theory, which is invariant under $SL(2,\R)$.  However this symmetry is broken by quantum corrections that generate an anomaly in the $U(1)$ R-symmetry that is consistent with the breaking of $SL(2,\R)$ to $SL(2,\Z)$ \cite{Gaberdiel:1998ui}, which is the duality symmetry of type IIB superstring theory.  The low-energy expansion (\ref{eq:LLeff}) is therefore invariant under $SL(2,\Z)$, and $\tau$ parameterises a fundamental domain that may be chosen to be $\mathcal{F}_\tau = \{ |\tau_1|\leq \frac{1}{2} \,,\,|\tau| \geq 1\}$.

 The second term  in (\ref{eq:LLeff}) is proportional to  $R^4$ \cite{Gross:1986iv, Grisaru:1986dk}, which is a specific contraction of four Riemann tensors that has mass dimension $8$ and is $1/2$-BPS, which means it preserves $16$ of the $32$ supersymmetries associated with ten-dimensional maximal supersymmetry.  
 Similarly, the higher-derivative term $d^4R^4$ has mass dimension $12$ and is $1/4$-BPS while $d^6R^4$ has mass dimension $14$ and and is $1/8$-BPS. Maximal supersymmetry determines  the Lorentz contractions of the tensor indices. It also forbids the presence of $R^2$ and  $R^3$ interactions. The  ellipsis  in (\ref{eq:LLeff}) represents the suppressed supersymmetric completion,  as well as higher-order terms and terms that contribute to $n$-point amplitudes with $n>4$,  which we will come back to later.    
 
 The four BPS terms displayed in  (\ref{eq:LLeff}) are those that contribute to the four-point amplitude and are protected by supersymmetry.  As we will see, their coefficients are modular functions that are  solutions to  specific Laplace equations that are determined by supersymmetry combined with $SL(2,\Z)$ invariance. 
 Some properties of the modular  functions  of relevance to this article are given in  \ref{sec:math}.  The function $E(s; \tau, \bar{\tau})$ is a non-holomorphic Eisenstein series, which satisfies the Laplace eigenvalue equation (\ref{eq:Es}). In its zero Fourier mode for $\tau_1$, this function has two terms that are power-behaved in $\tau_2=1/g_s$ which are interpreted as perturbative contributions.  The  two perturbative contributions to the coefficient of the   $R^4$ interaction, $g_s^{-\frac{1}{2}}E(\frac{3}{2}; \tau, \bar{\tau})$   \cite{Green:1997tv, Green:1997as,Green:1998by}, correspond to tree-level and genus-1 ($\tau_2^2$ and $\tau_2^0$) contributions. 
 
 Similarly the coefficient of $d^4R^4$ is $g_s^{\half} \, E(\fiveh;\tau,\bar\tau)$  \cite{Green:1999pu}, which has perturbative terms corresponding to tree-level and genus-2 ($\tau_2^2$ and $\tau_2^{-2}$), but no genus-1 contribution. The absence of higher order perturbative terms implies that $R^4$ gets no contribution beyond genus 1  and $d^4 R^4$ gets no contribution beyond genus 2 in string perturbation theory. 
 The coefficient of $d^6R^4$ is the generalised non-holomorphic  Eisenstein series   $ g_s\,\mathcal{E}(\threeh, \threeh; 3; \tau, \bar{\tau})$ \cite{Green:2005ba}, which satisfies the inhomogeneous Laplace eigenvalue equation,  (\ref{eq:generalise}).   In this case its zero Fourier mode has four power-behaved terms that correspond to contributions from genus-$0$ up to genus-$3$  in superstring perturbation theory, and no higher-order perturbative terms.

There are many other fractional BPS terms in the effective action that have not been explicitly displayed in (\ref{eq:LLeff}). Many of these can be obtained by considering the low-energy expansion of  $n$-point amplitudes with $n\ge 5$.
In order to describe such amplitudes it is important to recall that the fluctuations of the  massless fields of type IIB supergravity around their background values carry specific $U(1)$ charges  \cite{Schwarz:1983qr,Howe:1983sra}.   The field $\tau$ has a non-zero background value, $\tau=\tau^0$, which defines the string coupling constant.  However, its fluctuation $Z$, defined  by the Cayley transformation
\bea
\label{teautoB}
Z= \frac{\tau - \tau^0}{\tau-\bar\tau^0}\,,  
\eea
carries $U(1)$ charge $-2$, while its conjugate $\bar Z$ has $U(1)$ charge $+2$ \cite{Green:2019rhz}.   A scattering amplitude is defined with a specified value of  the string coupling ${\rm Im}(\tau_0)=1/g_s$ and a $U(1)$ transformation is identified with a  $SL(2,\Z)$ transformation that leaves the background value $\tau_0$ unchanged. The 256 physical states in the type IIB supergravity supermultiplet have $U(1)$ charges ranging from $-2$ to $+2$ in our conventions.     The total $U(1)$ charge violation of a $n$-point amplitude  with massless external states is generally non-zero and satisfies the inequality $|q_{U(1)}| \le 2(n-4)$.  It follows that the  $U(1)$ charge is conserved in all four-point functions but may be violated when $n>4$.

 One particular example of a well-studied amplitude that violates $U(1)$  is the sixteen-dilatino interaction. The dilatino, $\Lambda$, carries $U(1)$ charge $q_{U(1)} = -3/2$ 
 so this interaction violates the $U(1)$ charge by $q_{U(1)}=-24$.  Using a M-theory duality  argument the  leading term in the low-energy limit was found \cite{Green:1997me} to be  proportional to  $g_s^{-\half}  E_{12}(\threeh;\tau,\bar\tau)  \Lambda^{16}$  where $E_w(s;\tau,\bar\tau)$ is a modular form with weight $(w,-w)$ that is defined in \eqref{eq:lattice}by acting on $E(s;\tau,\bar\tau) $ with $w$ modular covariant derivatives.\footnote{ Some relevant properties of modular forms  are briefly described  in \ref{sec:math}. They have holomorphic and anti-holomorphic modular weights  $(w,-w)$.} This  expression was also produced by an argument based directly on the supersymmetry transformations of the fields in type IIB supergravity     \cite{Green:1998by}.  We will shortly demonstrate that this argument can be simplified and generalised by the use of a ten-dimensional spinor-helicity  superspace formalism.  
 
  The $\Lambda^{16}$ amplitude is one example of a maximally $U(1)$-violating (MUV) amplitude  \cite{Boels:2012zr}, which violates $U(1)$ by precisely $-2(n-4)$ units.\footnote{Note that a `minimally $U(1)$-violating amplitude' violates the $U(1)$ charge by  $q_{U(1)}=2(n-4)$ units. This sign convention ensures that the coefficient function multiplying a maximally/minimally $U(1)$-violating amplitude has maximal/minimal holomorphic weight $w=\pm(n-4)$.} For all such amplitudes  the terms up to mass dimension $14$ in the low-energy expansion are  BPS terms and their coefficients  are modular forms that were determined  by supersymmetry in  \cite{Green:2019rhz} using superamplitude methods that we will now describe.   The procedure uses the fact BPS coefficients arising in the low-energy expansion of any MUV $n$-point amplitude  are related by supersymmetry to the coefficients in the low-energy expansion of the amplitude  for four gravitons, denoted by $h$, and $(n-4)$ complex scalars $Z$, $\langle h\,h\,h\,h\,  \underbrace {Z\,\dots\, Z \rangle}_{n-4}$.
  
  These amplitudes give rise to BPS interactions that have the symbolic form  
\bea
(\alpha^{\prime})^{-1} R^4 Z^{n-4}\, , \qquad \alpha^{\prime} d^4R^4 Z^{n-4} \, , \qquad (\alpha^{\prime})^2 d^6R^4 Z^{n-4}\, .
\eea
Maximal supersymmetry ensures that there is a unique Lorentz scalar for $R^4 Z^{n-4}$ and $d^4R^4Z^{n-4}$, respectively. However, as we will explain later making use of  superamplitude methods, there are two and only two independent Lorentz scalars that contribute tor $d^6R^4 Z^{n-4}$ when $n \geq 6$. 

In the next subsection we will show that the coefficient of $(\alpha^{\prime})^{-1} R^4 Z^{n-4}$ is proportional to $E_{n-4}(\threeh; \tau,\bar{\tau})$, and the coefficient of $\alpha^{\prime} d^4R^4 Z^{n-4}$ is proportional to $E_{n-4}(\fiveh; \tau,\bar{\tau})$.  In the case of $d^6R^4 Z^{n-4}$  there are two  invariant tensor structures when $n\geq 6$.  The coefficient associated with one of these is simply $\mathcal{E}_{n-4}(\threeh, \threeh; 3; \tau,\bar{\tau})$ defined in (\ref{MEw}), but the coefficient of the other structure is a new modular form, which will be described  in the following.

\subsection{Superamplitudes and low-energy expansion}
\label{eq:superamp}

The methods we will use for studying these higher-derivative terms were first introduced in \cite{Wang:2015jna}, which applied modern amplitude techniques to rederive the results of \cite{Green:1997tv, Green:1997as,Green:1998by, Green:1999pu, Green:2005ba} for the four-point interactions that are explicitly displayed in \eqref{eq:LLeff}. We will follow closely the discussion given in  \cite{Green:2019rhz}, which treats $n$-point MUV interactions uniformly, with the four-point interactions as a special case by setting $n=4$.  It proves very useful to introduce a ten-dimensional spinor-helicity formalism \cite{Boels:2012ie}, which is the analogue of the more familiar four-dimensional formalism.   This expresses the momentum $k_\mu$ of any massless state in ten dimensions in terms of chiral  bosonic spinors $\lambda^A_a$,
\bea
k^{BA}:= (\gamma^{\mu})^{BA}\, k_\mu = \lambda^{Ba} \lambda_a^{A}\, ,
\label{momspin}
\eea
where $A=1,\dots,16$  labels  the components of a $SO(9,1)$ chiral spinor, $a=1,\dots ,8$ labels the components of a $SO(8)$ spinor of the  little group of massless states, and  $( \gamma^{\mu})^B  _{\  A}$ are ten-dimensional gamma matrices. The Grassmann variables $\eta^a$ encode type IIB supersymmetry, where the supercharges are expressed as \cite{Boels:2012ie}
\bea
q^A_i=\lambda^A_{i,a}\, \eta^a_i \, ,\qquad\quad \bar q^B_i= \lambda^{B,a}_i\, \frac{\partial}{\partial \eta^a_i}\,,
\label{susyc}
\eea
satisfying the on-shell super-algebra  
\bea
\{\bar q^B_i,\, q^A_i \} =    \lambda^{Ba}_i \lambda_{i, a}^{A} = p_i^{BA}\, ,
\label{supalg}
\eea 
and the index $i=1, 2, \ldots ,n$ labels the $n$ particles scattered.

The massless physical states are packaged into a superfield that has the following expansion in powers of  $\eta_i$ 
\bea 
\Phi (\eta_i)= Z+\eta_i^a  \Lambda_a + \frac{1}{2!} \eta_i^a\eta_i^b \phi_{ab}  +\dots+\frac{1}{8!} (\eta_i)^8 \bar {Z} \,.
\label{phiexp}
\eea
 The superfield $\Phi (\eta_i)$ is assigned a $U(1)$ charge $q_\Phi=-2$, and $\eta$ is assigned $U(1)$ charge $q_\eta= - 1/2$. Therefore a component field with $m$  $SO(8)$ spinor indices  has a charge $q_m= -2+m/2$.  For instance, the scalar field $Z$ has charge $-2$, and the graviton $h$, has $U(1)$ charge $0$.

A $n$-point superamplitude then is a function of $\lambda_i, \eta_i$, with $i=1,2,\cdots, n$, and supersymmetry implies that the superamplitude should take the form, 
  \bea
 A_n(\lambda_i, \eta_i) =\delta^{10}\left(\sum_{i=1}^n k_i\right)\, \delta^{16}(Q_n)\, \hat A_n(\lambda_i, \eta_i)\,,
  \quad {\rm with} \quad \bar Q_n^B\hat A_n=0\, ,
  \label{ahatprop}
  \eea
where $Q^B_n = \sum_{i=1}^n q^B_i$ and $\bar Q^B_n =\sum_{i=1}^n  \bar q^B_i$ (with $B=1,2,\cdots, 16$) are the total supercharges. The formula (\ref{ahatprop}) ensures that the superamplitude $A_n$ is annihilated by the thirty-two supersymmetries.

Apart from the three-particle on-shell amplitude, which has degenerate kinematics, these conditions imply that scattering amplitudes vanish unless the total number of $\eta$'s from external states is at least $16$. 
Amplitudes in which there are exactly sixteen $\eta$ variables are those for which $q_U=- 2(n-4)$ --  they are MUV amplitudes.    
 In this case the quantity $\hat A_n$  contains no factors of $\eta$.  Therefore it is a function  of the Mandelstam variables, $s_{ij} = - \alpha'/4\, (k_i + k_j)^2$, that encodes the $\alpha'$-dependence characteristic of string theory, as well  as the dependence  on the complex coupling constant,  $\tau$.

In considering the low-energy expansion of amplitudes it is important to take into  account non-analytic features  that come from the effects of higher genus contributions and non-perturbative effects.  Although this is very complicated in  general, 
the  first three terms in the low-energy expansion  of  the ten-dimensional  amplitude, which are protected by supersymmetry, are analytic in the Mandelstam invariants.  These terms correspond to the first three terms of the $\alpha'$ expansion  of $\hat A_n$, which are symmetric polynomials of degree $p=0$,  $p=2$ and $p=3$ in the Mandelstam invariants, since maximal $U(1)$-violating amplitudes cannot have poles in momenta and the case $p=1$ vanishes identically $\sum_{i<j} s_{ij}=0 $.

This consideration leads to BPS terms  in the low-energy limits of $n$-particle  superstring amplitudes in the form,\footnote{The amplitude is defined in a given background $\tau=\tau^0$. Here and in what follows, to simplify the notation we will drop the superscript $0$ of the background field (or equivalently the coupling) $\tau^0$ and denote it by $\tau$.} 
 \bea    
 \label{eq:maximal-U(1)}
 A_n^{(p)} (\lambda_i, \eta_i )=  F^{(p)}_{n-4} (\tau,\bar{\tau})\, \delta^{16}(Q_n) \, \hat{A}^{(p)}_n(s_{ij}) \,,
  \eea
where the subscript $(n-4)$ indicates the $U(1)$ weight, $w=n-4$.  In this expression, which includes amplitudes of the form  (\ref{ahatprop}),   the factor  $\hat{A}^{(p)}_n(s_{ij})$ is simply a symmetric homogeneous degree-$p$ polynomial of Mandelstam invariants. The case $p=0$,  
\bea
\hat{A}^{(0)}_n(s_{ij}) =1\, , 
\eea 
is identified with  $R^4Z^{n-4}$ and its supersymmetric completion in the effective Lagrangian. Note that $\delta^{16}(Q_n)$ has power counting as $\delta^{16}(Q_n)\sim (\lambda)^{16} \sim k^8$, which indeed has $8$ derivatives, just as $R^4Z^{n-4}$.   $A_n^{(2)}$ is identified with  $d^4R^4Z^{n-4}$ and its supersymmetric completion, with
\bea
\hat{A}^{(2)}_n(s_{ij}) \equiv \mathcal{O}^{(2)}_{n}(s_{ij}) = {1\over 2} \sum_{i<j} s_{ij}^2  \, .
\eea 
Finally, $A_n^{(3)}$ is identified with  $d^6R^4Z^{n-4}$ and its supersymmetric completion. As anticipated earlier, there are two independent structures at the order $d^6R^4Z^{n-4}$ when $n\geq 6$, which the superamplitude description makes very explicit. This follows from the fact that there are two independent degree-3 symmetric polynomials,
\bea \label{eq:p3}
\hat{A}^{(3)}_{n, 1}(s_{ij}) = \sum_{i<j} s_{ij}^3 \, ,\qquad {\rm or } \qquad \hat{A}^{(3)}_{n,2}(s_{ij}) = \sum_{i<j<k} s_{ijk}^3 \, ,
\eea
with $s_{ijk} = -\alpha'/4 \, (k_i+k_j+k_k)^2$, each of which is associated with  a coupling-dependent coefficient $F^{(3)}_{n-4, r} (\tau,\bar{\tau})$ for $r=1,2$. 
Note that for $n=4$ we have $\hat{A}^{(3)}_{4,2}(s_{ij}) =0$, while for $n=5$ $\hat{A}^{(3)}_{5,2}(s_{ij})  = \hat{A}^{(3)}_{5,1}(s_{ij})$, so that $\hat{A}^{(3)}_{n,2}(s_{ij}) $ only plays a r\^ole from $n=6$ onwards.

The coefficient functions $F^{(p)}_{n-4} (\tau,\bar{\tau})$ with $p=0,2,3$ contain the full non-perturbative dependence on the complex type IIB coupling  constant. When $n=4$, these are the modular functions reviewed in the previous section, so that $F^{(0)}_{0} (\tau,\bar{\tau}) \propto E(\threeh; \tau,\bar{\tau})$, $F^{(2)}_{0} (\tau,\bar{\tau}) \propto E(\fiveh; \tau,\bar{\tau})$, and $F^{(3)}_{0} (\tau,\bar{\tau}) \propto \mathcal{E}(\threeh, \threeh; 3; \tau,\bar{\tau})$.  

Terms of higher order in the low-energy expansion -- i.e. of mass dimension $\ge 16$  (or $p \ge 4$) -- are $D$-terms and they can be written in terms of a function $f(\lambda,\eta)$ multiplied by all $32$ supercharges.   For example  if $\hat A_n^{(4)}(s_{ij})$ is a symmetric polynomial in Mandelstam invariants of degree $4$ it can be expressed in the schematic form
\bea \label{eq:nonBPS}
\hat A_n^{(4)}(s_{ij}) \sim \sum_{\rm permutations} (\bar Q)^{16} \eta_i^8  \eta_j^8\, .
\eea
This is simply a consequence of power counting since $(\bar Q)^{16}$ is of order $s_{ij}^4$. By construction, $\hat A_n^{(4)}$ given above is annihilated by all $16$ $\bar Q$'s. As we will see later such terms are,  unsurprisingly, unconstrained and do not appear to be protected by supersymmetry.

\subsection{Soft-dilaton constraints} 

The  behaviour of amplitudes in limits in which one or more of the momenta of the scattering particles is zero  (soft limits)   is intimately related to symmetry properties of the theory. A prototype  is the Adler zero \cite{Adler:1964um}, which refers to the vanishing of amplitudes  for scattering of Goldstone bosons in the chiral non-linear sigma model as one of the momenta is taken to be soft.  Similarly, taking the soft limit  of a dilaton $Z$, with momentum $p_n$, in type IIB supergravity gives
\bea \label{eq:softsugra}
A^{SG}_n (X, Z(k_n)) \Big{|}_{k_n \rightarrow 0} = 0 \,  , 
\eea
where $SG$ indicates a supergravity amplitude and $X$ denotes the remaining $(n-1)$ scattered fields. This soft behaviour reflects the coset structure $SL(2, \mathbb{R})/U(1)$ of type IIB supergravity. However, stringy effects break the $U(1)$ symmetry and the soft-dilaton limit of string amplitudes no longer vanishes.\footnote{Since $Z$ is a combination of axion and dilation, we  are really considering both the axion and the dilation to be soft, even though we are calling this a  soft-dilaton condition.} The result is \cite{Green:2019rhz},  
\bea  \label{eq:softstring}
\!\!\! \!\!\! \!\!\! \!\!\! \!\!\! \!\!\!  \!\!\! \!\!\! \!\!\! A_n (X, Z(k_n)) \Big{|}_{k_n \rightarrow 0} = 2 \cD_{w} A_{n-1} (X) \, , \quad A_n (X, \bar{Z}(k_n)) \Big{|}_{k_n \rightarrow 0} = 2  \bar \cD_{-w} A_{n-1} (X) \, ,
\eea
where $w$ is the $U(1)$ weight of the lower-point amplitude $A_{n-1} (X)$. In  supergravity, all the amplitudes have zero $U(1)$ weights, therefore the above soft dilaton relations reduce to (\ref{eq:softsugra}). Furthermore, one may consider the sum of the two soft dilaton relations in (\ref{eq:softstring}), which projects out the Ramond--Ramond pseudoscalar (i.e. the axion field) and  leads to  a soft relation for the real dilaton  \cite{Ademollo:1975pf, Shapiro:1975cz, DiVecchia:2015jaq},
\bea
A_n (X, Z(k_n) +  \bar{Z}(k_n) ) \Big{|}_{k_n \rightarrow 0} = 2\left( \cD_{w} + \bar \cD_{-w} \right)A_{n-1} (X) \, .
\eea
Applying the relations (\ref{eq:softstring}) to the low-energy expansion of MUV amplitudes given in (\ref{eq:maximal-U(1)}), we find
\bea
\!\!\!\!\!\!\!\!F^{(p)}_{n-4} (\tau,\bar{\tau})\, \delta^{16}(Q_n) \, \hat{A}^{(p)}_n(s_{ij})  \Big{|}_{k_n \rightarrow 0}  = 2 \cD_{n-5} F^{(p)}_{n-5} (\tau,\bar{\tau})\, \delta^{16}(Q_{n-1}) \, \hat{A}^{(p)}_{n-1}(s_{ij}) \, ,
\eea
where we have used the fact $w$ in \eqref{eq:softstring} is $n-5$ for the MUV amplitudes. 
Since $Z$ is the top component of the on-shell superfield (\ref{phiexp}) without any  $\eta$  factors, we see  that $\delta^{16}(Q_n)$ reduces to $\delta^{16}(Q_{n-1})$ directly in the soft-dilaton limit, therefore, 
\bea
F^{(p)}_{n-4} (\tau,\bar{\tau})\,  \hat{A}^{(p)}_n(s_{ij})  \Big{|}_{k_n \rightarrow 0}  = 2 \cD_{n-5} F^{(p)}_{n-5} (\tau,\bar{\tau})\, \hat{A}^{(p)}_{n-1}(s_{ij}) \, .
\eea

We will discuss this relation for values of $p\le 3$.  In the case $p=0$, $\hat{A}^{(0)}_n(s_{ij})  =1$, so the soft limit is trivial and we obtain
\bea \label{eq:soft1}
F^{(0)}_{n-4} (\tau,\bar{\tau})  = 2 \cD_{n-5} F^{(0)}_{n-5} (\tau,\bar{\tau})  \, .
\eea
In the case of $p=2$, it is easy to see $\hat{A}^{(2)}_n(s_{ij}) \Big{|}_{k_n \rightarrow 0} =\hat{A}^{(2)}_{n-1}(s_{ij})$, we again obtain
\bea \label{eq:soft2}
F^{(2)}_{n-4} (\tau,\bar{\tau})  = 2 \cD_{n-5} F^{(2)}_{n-5} (\tau,\bar{\tau})  \, .
\eea
Therefore, recalling that  $F^{(0)}_{0} (\tau,\bar{\tau}) = E(\threeh; \tau,\bar{\tau})$ and  $F^{(2)}_{0} (\tau,\bar{\tau})= E(\fiveh; \tau,\bar{\tau})$,  the above relations  uniquely determine $F^{(0)}_{n-4} (\tau,\bar{\tau})$ and $F^{(2)}_{n-4} (\tau,\bar{\tau})$ for any $n$. 

As shown in (\ref{eq:p3}), the story becomes more interesting for $p=3$, in which case there are two independent polynomials when $n\geq 6$. As was argued in \cite{Green:2019rhz}, it is important to choose particular linear combinations of $\hat{A}^{(3)}_{n, 1}(s_{ij})$ and $\hat{A}^{(3)}_{n, 2}(s_{ij})$ to form the basis for  the amplitude $ A_n^{(3)}$. In particular, we choose, 
\bea \label{eq:basis}
\mathcal{O}^{(3)}_{n,1} &= {1\over 32} \left[ (28-3n)\hat{A}^{(3)}_{n, 1}(s_{ij}) + 3 \hat{A}^{(3)}_{n, 2}(s_{ij})\right] \, , \cr
\mathcal{O}^{(3)}_{n,2} &=(n-4)\hat{A}^{(3)}_{n, 1}(s_{ij}) - \hat{A}^{(3)}_{n, 2}(s_{ij}) \, ,
\eea
so the amplitude is given by
\bea \label{eq:p=3}
A_n^{(3)} =\delta^{16}(Q_n) \left[ F^{(3)}_{n-4, 1} (\tau,\bar{\tau}) \mathcal{O}^{(3)}_{n,1} (s_{ij}) +  F^{(3)}_{n-4, 2} (\tau,\bar{\tau}) \mathcal{O}^{(3)}_{n,2} (s_{ij})\right]\,.
\eea
With this particular linear combination, when $n=6$ the term involving 
\begin{equation}
\mathcal{O}^{(3)}_{6,1}= \frac{1}{32}\Big(10\!\! \sum_{1\leq i{<}j\leq6} s_{ij}^3+ 3\!\!\!\sum_{1\leq i<j<k\leq 6} s_{ijk}^3\Big)\,,
\end{equation} is identified with the term of mass dimension $14$  in the low-energy expansion of the six-point MUV tree-level amplitude from explicit computation as given in \eqref{eq:tree}.  Hence the coefficient $F^{(3)}_{2, 2} (\tau,\bar{\tau})$ of the second linear combination $\mathcal{O}^{(3)}_{6,2}$ does not receive any tree-level contribution but only contains terms originating from higher-genus string amplitudes.  Furthermore, $\mathcal{O}^{(3)}_{n,2}$ vanishes for $n=4$ and $n=5$, while for $n=6$, $\mathcal{O}^{(3)}_{6,2} \sim \sum_{\rm perm} s_{12} s_{34} s_{56}$ vanishes in the soft limit, which has important consequences as we will discuss later.

The preceding argument leads to expressions for  $\mathcal{O}^{(3)}_{n,1}$ and  $\mathcal{O}^{(3)}_{n,2}$  for all values of  $n$. They are   determined uniquely by the following soft limits, 
\bea
\mathcal{O}^{(3)}_{n,1}(s_{ij}) \Big{|}_{k_n \rightarrow 0} = \mathcal{O}^{(3)}_{n-1,1}(s_{ij}) \, , \qquad \quad \mathcal{O}^{(3)}_{n,2}(s_{ij}) \Big{|}_{k_n \rightarrow 0} = \mathcal{O}^{(3)}_{n-1,2} (s_{ij})\, ,
\eea
which give 
\bea
  \mathcal{O}^{(3)}_{n,1}(s_{ij}) &= \frac{1}{32} \left[  (28-3n) \sum_{i<j} s^3_{ij} +3 \sum_{i<j<k} s^3_{ijk} \right]  \, , \cr
  \mathcal{O}^{(3)}_{n,2}(s_{ij}) &= (n-4) \sum_{i<j} s^3_{ij} - \sum_{i<j<k} s^3_{ijk} \, .
\eea
These properties and the soft-dilaton conditions imply that the coefficients $F^{(3)}_{n-4, 1} (\tau,\bar{\tau}),$ and $F^{(3)}_{n-4, 2} (\tau,\bar{\tau})$ obey the following relations, 
\bea \label{eq:d6R4}
\!\!\!\!\!\!\!F^{(3)}_{n-4, 1} (\tau,\bar{\tau}) = 2 \cD_{n-5} F^{(3)}_{n-5, 1} (\tau,\bar{\tau}) \, , \qquad  F^{(3)}_{n-4, 2} (\tau,\bar{\tau})  = 2 \cD_{n-5} F^{(3)}_{n-5, 2} (\tau,\bar{\tau}) \, .
\eea
Importantly, the second equation only applies to the cases with $n>6$. Therefore, $F^{(3)}_{n-4, 1} (\tau,\bar{\tau})$ is determined recursively by $F^{(3)}_{0, 1} (\tau,\bar{\tau})$, which is  the coefficient of $d^6R^4$ that is  given by the generalised non-holomorphic Eisenstein series $\mathcal{E}(\threeh, \threeh;3;\tau,\bar{\tau})$.  The other coefficient $F^{(3)}_{n-4, 2} (\tau,\bar{\tau})$ is new, and  will be determined separately. 

It is instructive to have explicit results of the low-energy expansion of the relevant string amplitudes. At  tree-level, these are relatively easy to determine. For example, the coefficients of the low-energy expansion of, tree-level  MUV amplitudes with up to six particles and up to $14$ derivatives are given by \cite{Green:2019rhz}
\bea \label{eq:tree}
\hat A_4(s_{ij}) &=& 2\tau_2^{3\over 2} \zeta(3)  + \tau_2^{5\over 2} \zeta(5)  \mathcal{O}^{(2)}_{4}(s_{ij})
+ {2\over 3} \tau_2^3 \zeta(3)^2 \mathcal{O}^{(3)}_{4,1}(s_{ij}) \, , \cr
\hat A_5(s_{ij}) &=& 3\tau_2^{3\over 2} \zeta(3)  + {5 \over 2} \tau_2^{5\over 2} \zeta(5)  \mathcal{O}^{(2)}_{5}(s_{ij})
+ 2 \tau_2^3 \zeta(3)^2 \mathcal{O}^{(3)}_{5,1}(s_{ij}) \, , \cr
\hat A_6(s_{ij}) &=& {15\over 2}\tau_2^{3\over 2} \zeta(3)  + {35 \over 4} \tau_2^{5\over 2} \zeta(5)  \mathcal{O}^{(2)}_{6}(s_{ij})
+ 8 \tau_2^3 \zeta(3)^2 \mathcal{O}^{(3)}_{6,1}(s_{ij}) \, . 
\eea
These tree-level results provide useful data for determining parameters in  the differential equations arising from supersymmetry constraints that will be discussed in the next section.
\subsection{Superamplitude constraints and differential equations}

We have seen that the soft-dilaton constraints relate coefficients  in the low-energy expansion of MUV amplitudes since $F^{(p)}_{w+1} (\tau,\bar{\tau}) \sim \cD_w F^{(p)}_{w} (\tau,\bar{\tau})$. Following \cite{Green:2019rhz}, we will now show that conjugate first order differential equations involving $\bar \cD$ are determined by supersymmetry constraints generalising the procedure of \cite{Wang:2015jna, Lin:2015ixa, Wang:2015aua, Chen:2015hpa, Bianchi:2016viy}. The key ingredient in this procedure, which has been checked in many examples, is that supersymmetric contact terms of mass dimension $\leq 14$ are not allowed for non-maximal $U(1)$-violating processes.\footnote{If there are more than $14$ derivatives, one can then construct supersymmetric contact terms. One of such examples is given in \eqref{eq:nonBPS}. } This fact implies that the low-energy expansion of a sueramplitude up to mass dimension $14$ is uniquely determined by lower-point amplitudes via factorisation using tree-level unitarity. 
In this section we will simply denote by $\cD$ and $\bar \cD$ the action of the holomorphic and anti-holomorphic covariant derivatives $\cD_w,\bar\cD_{-w}$ (given in \eqref{covderdef}) on a modular function with weight $(w,-w)$ as to avoid cluttering the notation. Given the fact that all of the coefficients $F^{(p)}_{n-4}$ have modular weights $(w,-w)$, with $w=n-4$ it should be clear which specific covariant derivatives $\cD,\bar\cD$ are acting on them. 

\begin{figure}
  \begin{center}
    \scalebox{0.85}{\includegraphics{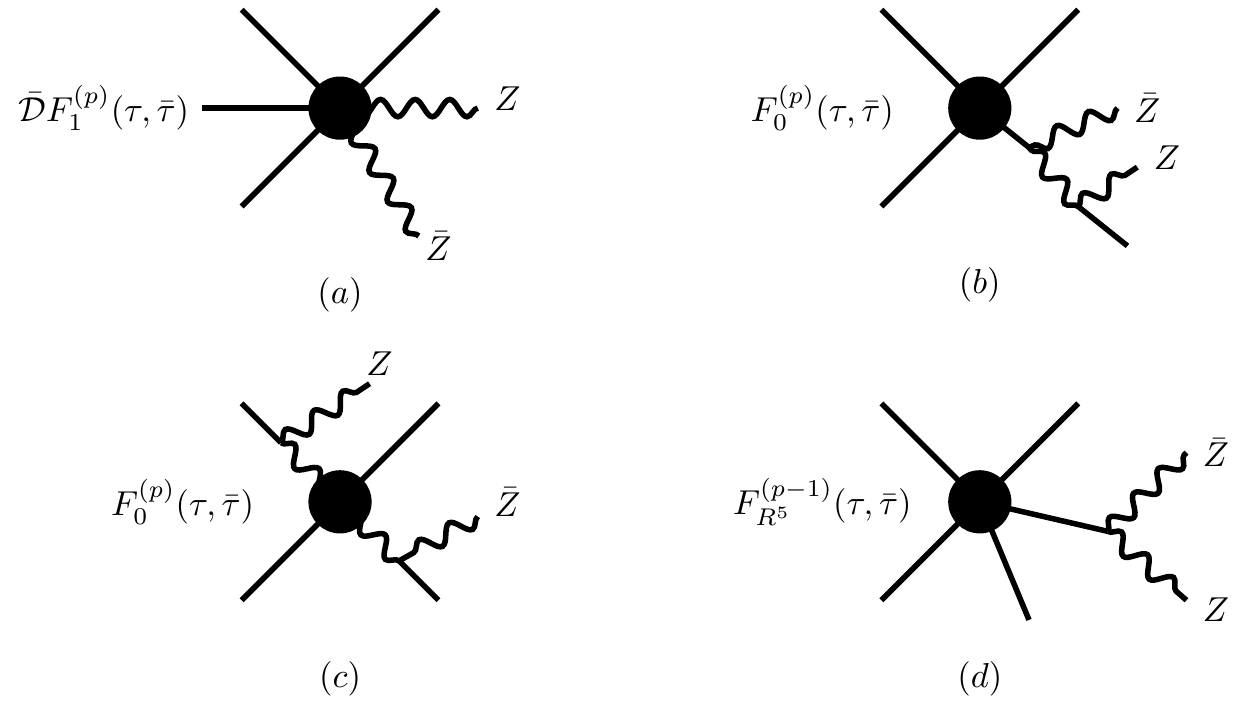}}
    \caption{The diagrams that contribute to the six-point amplitude $A_6(h,h,h,h, Z, \bar{Z})$ at order $R^4$ or $d^4R^4$, with $h$ being the graviton (represented by the straight lines) and $Z, \bar{Z}$ the axion-dilaton field and its conjugate (represented by the wavy lines). In (a) $F_1^{(p)}(\tau,\bar{\tau})$ is the coefficient of the interaction $d^{2p}R^4Z$. In (b) $F_0^{(p)}(\tau,\bar{\tau})$ is the coefficient of the interaction $d^{2p}R^4$ (and its supersymmetric relative in (c)).  In (d) $F_{R^5}^{(p-1)}(\tau,\bar{\tau})$ is the coefficient of the interaction $d^{2p-2}R^5$. }
    \label{fig:6ptsamp}
  \end{center}
\end{figure}

\subsubsection*{\underline{Cases $p=0$ and $p=2$.}}

We will illustrate the idea by considering a six-point amplitude with four gravitons, together with a  $Z$ and a $\bar Z$, which is a $U(1)$-conserving amplitude.  The diagrams that contribute to this amplitude are sketched in Fig.~\ref{fig:6ptsamp},  which contains three factorisation diagrams and one contact diagram.  
As emphasised earlier,  the contact term alone is inconsistent with supersymmetry, 
therefore it must be linearly related to the factorisation diagrams.  In other words,  the absence of a non-MUV supersymmetric contact term implies a linear relation among all these terms, and consequently a linear relation of the corresponding coefficients, 
\bea \label{eq:cD1}
\bar \cD F^{(p)}_1(\tau,\bar{\tau}) + a_1 \, F^{(p)}_0(\tau,\bar{\tau})  + a_2 \, F_{R^5}^{(p-1)}(\tau,\bar{\tau}) =0\, ,
\eea
where the coefficients $a_1$ and $a_2$ are in principle computable by  explicit evaluation of  the contributions in figure~\ref{fig:6ptsamp}.   However, that is very complicated and so in the following  the coefficients will be fixed by comparison with string perturbation theory.  

Note, if $p=0$ (i.e. at order $R^4$), the contribution of the $R^5$ vertex in figure~\ref{fig:6ptsamp}-(d)  vanishes. If $p=2$, one can further relate  the $R^5$ coefficient, $F_{R^5}^{(p-1)}(\tau,\bar{\tau})$, to  $F^{(p)}_0(\tau,\bar{\tau})$ by considering the five-graviton amplitude, which is a non maximal $U(1)$-violating process. It receives contributions from $d^{2p-2}R^5$    (with coefficient $F_{R^5}^{(p-1)}(\tau,\bar{\tau})$) and pole terms arising from attaching a three-graviton vertex to $d^{2p}R^4$ (with coefficient  $F^{(p)}_0(\tau,\bar{\tau})$). By the same argument, this leads to a  linear relation between their coefficients
\bea   \label{eq:R5}
F^{(p-1)}_{R^5}(\tau,\bar{\tau}) + a_3  \,  F^{(p)}_{0}(\tau,\bar{\tau}) =0\, ,
\eea
that is  in agreement with \cite{Richards:2008jg}.  By combining (\ref{eq:cD1}) and (\ref{eq:R5}), we  arrive at 
\bea \label{eq:cD2}
\bar \cD F^{(p)}_1(\tau,\bar{\tau}) + a_4 \, F^{(p)}_0(\tau,\bar{\tau})  =0\,, 
\eea
which, together with (\ref{eq:soft1}) or (\ref{eq:soft2}), leads to the Laplace equation
\bea
\Delta_\tau F^{(p)}_0(\tau,\bar{\tau}) + 2 a_4 \, F^{(p)}_0(\tau,\bar{\tau})  =0 \,.
\eea
Although the constant $a_4$ is computable, in principle, this is not straightforward. As we commented earlier, however, it can also be determined from knowledge of the tree-level behaviour of the string amplitudes,  which implies $F^{(0)}_0(\tau,\bar{\tau}) \sim \tau_2^{{3\over 2}}$ and $F^{(2)}_0(\tau,\bar{\tau}) \sim \tau_2^{{5\over 2}}$. Therefore, 
\bea 
\left(\Delta- {3\over 4} \right) F^{(0)}_0(\tau,\bar{\tau})   =0 \, , \qquad \left(\Delta- {15\over 4} \right) F^{(2)}_0(\tau,\bar{\tau}) =0 \, ,
\eea
reproducing the Laplace equations for non-holomorphic Eisenstein series. We therefore find $F^{(0)}_0(\tau,\bar{\tau})= E(\threeh; \tau,\bar{\tau})$ and $F^{(2)}_0(\tau,\bar{\tau}) =\half E(\fiveh; \tau,\bar{\tau})$.   Once $F^{(p)}_0(\tau,\bar{\tau})$ is determined, (\ref{eq:soft1}) and  (\ref{eq:soft2}) fix all the $F^{(p)}_{n-4}(\tau,\bar{\tau})$ for any $n$.  These results have been confirmed by explicit  perturbative string theory calculations at genus-one and genus-two as described in the next section, as well as  by a  leading order D-instanton calculation in the case of $R^4$ term \cite{Sen:2021tpp, Sen:2021jbr}. 

\subsubsection*{\underline{Case $p=3$.}}  
\begin{figure}
  \begin{center}
    \scalebox{0.75}{\includegraphics{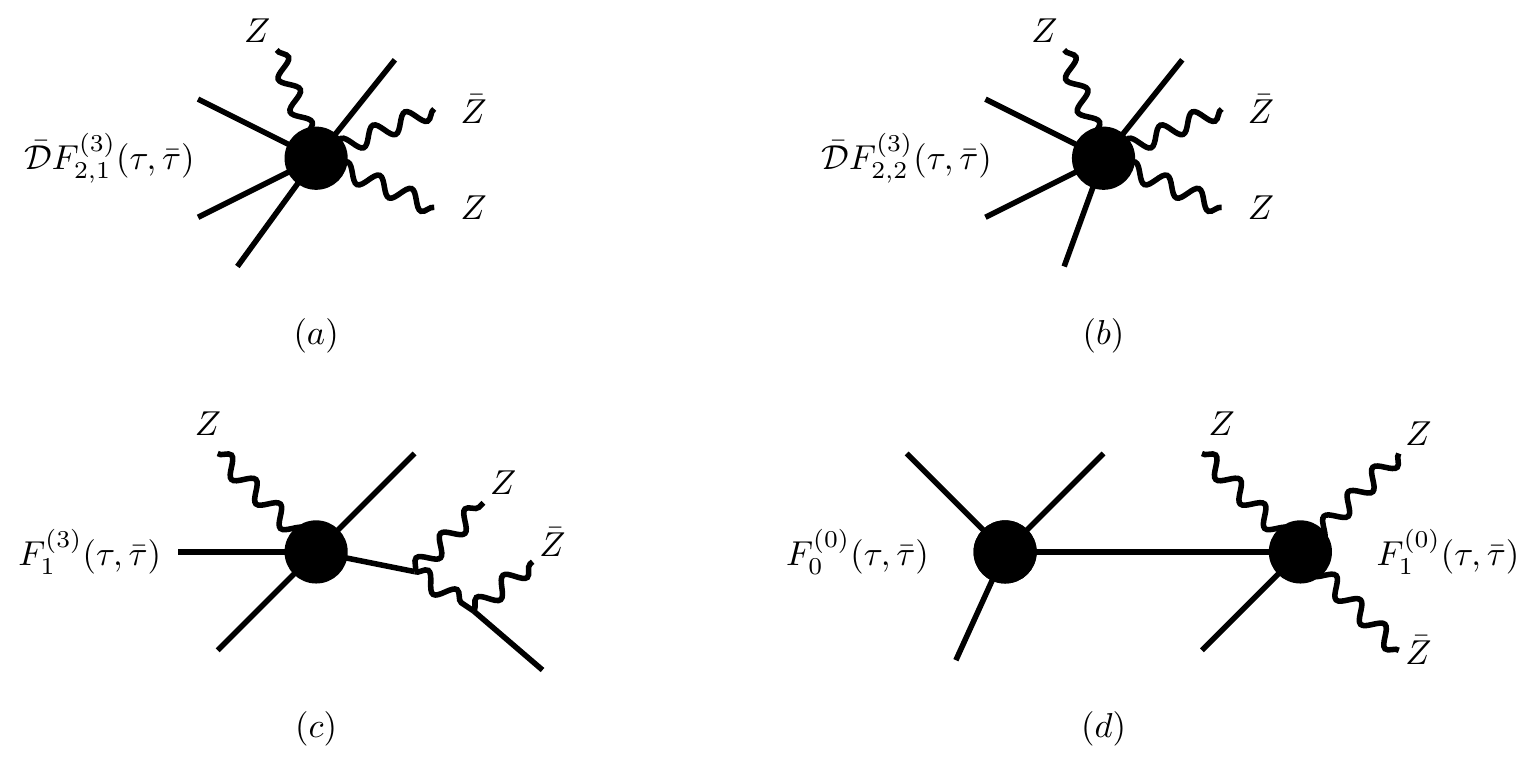}}
    \caption{The diagrams that contribute to the seven-point amplitude $A_7(h,h,h,h, Z, Z, \bar{Z})$ at order $d^6R^4$. }
    \label{fig:7ptsamp}
        \label{fig:sevenpart}
  \end{center}
\end{figure}

In the $p=3$ case additional diagrams contribute to the amplitude, see shown in figure~\ref{fig:7ptsamp}.   In addition to diagrams that are similar to those in figure~\ref{fig:6ptsamp},  a new  type of diagram arises consisting of  two  $p=0$ higher-derivative vertices  connected with a propagator shown in figure~\ref{fig:7ptsamp}-(d). The supersymmetry constraint  that implies the absence of contact terms leads to the relation
\bea
\bar \cD F^{(3)}_1(\tau,\bar{\tau})  +a \, F^{(3)}_0(\tau,\bar{\tau}) + b \, (F^{(0)}_0(\tau,\bar{\tau}))^2=0 \,,
\eea
where for $n<6$  $F^{(3)}_{n-4}(\tau,\bar{\tau}) \equiv  F^{(3)}_{n-4,1}(\tau,\bar{\tau})$ since, as  discussed below \eqref{eq:p=3}, the coefficient $F^{(3)}_{n-4,2}(\tau,\bar{\tau})$ makes its first appearance at $n=6$.
Combining this equation with the first equation in (\ref{eq:d6R4}) (since we are considering the case $n=4$  only the first equation in (\ref{eq:d6R4}) applies) leads to the Laplace equation for a generalised non-homomorphic Eisenstein series, after fixing the constants $a, b$ using perturbative superstring amplitude, 
\bea
\left(\Delta- 12 \right) F^{(3)}_0(\tau,\bar{\tau})   = -  (F^{(0)}_0(\tau,\bar{\tau}))^2\, .
\eea
Using the  fact that  $F^{(0)}_0(\tau,\bar{\tau}) = E(\threeh;\tau,\bar{\tau})$, we see that $F^{(3)}_0(\tau,\bar{\tau}) = \mathcal{E}(\threeh, \threeh; 3; \tau,\bar{\tau})$. We can then use the first equation in (\ref{eq:d6R4}) to determine all the coefficients $F^{(3)}_{n-4,1}(\tau,\bar{\tau})$ with $n>4$ associated with the higher-derivative terms $\mathcal{O}^{(3)}_{n-4,1}(s_{ij})$.

The constraints on the coefficient $F^{(3)}_{n-4,2}(\tau,\bar{\tau})$, which is associated with the higher-derivative term $\mathcal{O}^{(3)}_{n-4,2}(s_{ij})$, have to be determined separately.  These follow from the structure of the non-MUV seven-point amplitude with four gravitons, two $Z$’s and one $\bar Z$.  The contributions to the amplitude are shown in figure~\ref{fig:sevenpart}.  The absence of supersymmetric contact terms implies that the coefficient of each contact vertex (namely,  $\bar \cD F^{(3)}_{2,1}(\tau,\bar{\tau})$ and $\bar \cD F^{(3)}_{2,2}(\tau,\bar{\tau})$) is linearly related to the coefficients of the factorising terms.  Therefore we have the following differential equations
\bea \label{eq:DF31}
\bar \cD F^{(3)}_{2,1}(\tau,\bar{\tau}) + a_1 \, F_1^{(3)}(\tau,\bar{\tau}) + a_2 \, F_0^{(0)}(\tau,\bar{\tau}) F_1^{(0)}(\tau,\bar{\tau}) &=0 \, , 
\eea
and  
\bea \label{eq:DF32}
\bar \cD F^{(3)}_{2,2}(\tau,\bar{\tau}) + b_1 \, F_1^{(3)}(\tau,\bar{\tau}) + b_2 \, F_0^{(0)}(\tau,\bar{\tau}) F_1^{(0)}(\tau,\bar{\tau}) &=0\, .
\eea
Equation (\ref{eq:DF31}) involving $F^{(3)}_{2,1}(\tau,\bar{\tau})$ does not give new information, and is consistent with the result obtained earlier. Equation (\ref{eq:DF32})  determines the new modular function $F^{(3)}_{2,2}(\tau,\bar{\tau})$ that we anticipated earlier.  
Note that, by applying $\cD$ to \eqref{eq:DF31} and using \eqref{eq:d6R4}, we can obtain an inhomogeneous Laplace eigenvalue equation for $F^{(3)}_{2,1}(\tau,\bar{\tau})$
\begin{equation}\label{eq:LapF321}
(\Delta_{(-)} -10) F^{(3)}_{2,1}(\tau,\bar{\tau})  = -\frac{15}{2} \Big( E_0(\threeh;\tau,\bar\tau) E_2(\threeh,\tau,\bar{\tau}) +\frac{3}{5} E_1(\threeh,\tau,\bar{\tau})^2\Big)\,,
\end{equation}
with $\Delta_{(-)}$ the suitable Laplace operator \eqref{eq:Lap}, and where the specific values of the constants $a_1=-5,a_2=1$ were fixed in \cite{Green:2019rhz}.

From the construction of the higher-derivative terms, $\mathcal{O}^{(3)}_{n,1}$ and $ \mathcal{O}^{(3)}_{n,2}$, we know that $F^{(3)}_{2,2}(\tau,\bar{\tau})$ should vanish at tree-level, which allows us to fix the relative ratio between $b_1$ and $b_2$, leading to  
\bea \label{eq:DF322}
\bar \cD F^{(3)}_{2,2}(\tau,\bar{\tau}) +b_1 \left[ F_1^{(3)}(\tau,\bar{\tau}) -{1\over 3}  F_0^{(0)}(\tau,\bar{\tau}) F_1^{(0)}(\tau,\bar{\tau}) \right] =0 \, ,
\eea
where we have used the expressions of the perturbative  part of  $F_0^{(0)}(\tau,\bar{\tau}), F_1^{(0)}(\tau,\bar{\tau})$ and $F_1^{(3)}(\tau,\bar{\tau})$.   This leaves one undetermined constant, $b_1$, which can be fixed by a one-loop calculation of the six-point MUV amplitude to the same order as $d^6R^4$.\footnote{The constant $b_1$ has recently been determined to be $9/32$ by computing the one D-instanton contribution to the amplitude \cite{Agmon:2022vdj}.} 

Furthermore using \eqref{eq:DF31} together with $F_0^{(0)}(\tau,\bar{\tau}) F_1^{(0)}(\tau,\bar{\tau})  = 3 \bar\cD\big[(E_1(\threeh;\tau,\bar{\tau}))^2\big]$ we arrive at
\begin{equation}
 F^{(3)}_{2,2}(\tau,\bar{\tau}) = \frac{b_1}{5} \Big[F_{2,1}^{(3)}(\tau,\bar{\tau}) - 2E_1(\threeh;\tau,\bar{\tau})^2 \Big]\,,
\end{equation}
which, thanks to \eqref{eq:LapF321}, leads to an inhomogeneous Laplace eigenvalue equation for $F^{(3)}_{2,2}(\tau,\bar{\tau})$
\begin{equation}
(\Delta_{(-)} -10)  F^{(3)}_{2,2}(\tau,\bar{\tau}) = -\frac{5b_1}{2}\Big(E_0(\threeh;\tau,\bar{\tau})E_2(\threeh;\tau,\bar{\tau})-E_1(\threeh;\tau,\bar{\tau})E_1(\threeh;\tau,\bar{\tau})\Big)\,.
\end{equation}
The perturbative part of $F^{(3)}_{2,2}(\tau,\bar{\tau})$ takes the following form 
\bea
F^{(3)}_{2,2}(\tau,\bar{\tau}) =  {8 b_1 \over 3} \Big[ \zeta(2)\zeta(3) \tau_2 -{4\over 15}\zeta(2)^2 \tau_2^{-1}+{1\over 15}\zeta(6) \tau_2^{-3} \Big] +O(e^{-2\pi \tau_2}) \, .
\eea
The result predicts the precise ratios of the perturbative contribution to the higher-derivative term $\mathcal{O}^{(3)}_{6,2}(s_{ij})$ at genus one, two and three. Once $F^{(3)}_{2,2}(\tau,\bar{\tau})$ is determined, all higher-point coefficients $F^{(3)}_{n-4,2}(\tau,\bar{\tau})$ with $n>6$ are then also fixed by (\ref{eq:d6R4}).

\section{Exact results for integrated  correlators in $\mathcal{N}=4$ SYM}
\label{sec:correlator}

This section will review some recent exact results concerning correlation functions in $\mathcal{N}=4$ SYM \cite{Brink:1976bc} and their connections with superstring amplitudes.  The most studied  example of gauge/gravity duality is the holographic relationship between $\mathcal{N}=4$ SYM with $SU(N)$ gauge group  and type IIB superstring theory in $AdS_5 \times S^5$. 
According to this duality,  correlation functions in $\mathcal{N}=4$ SYM are the images of  scattering amplitudes of type IIB superstring theory.  In particular, the large-$N$ expansions of $\mathcal{N}=4$ correlators should reproduce the low-energy expansion of superstring amplitudes.  However, we will determine exact results that are valid for any finite value of $N$, as well as in the large-$N$ limit.  We will also study correlators of  $\mathcal{N}=4$ SYM with general classical gauge groups, and correlators of more than four operators that are dual to the MUV amplitudes that we met in the last section. 

Our emphasis will be on {\it integrated}  correlators of operators in the stress tensor supermultiplet.  That is, correlators of $1/2$-BPS operators that are integrated over the positions of  the operators with particular measures that are  chosen to preserve some of the supersymmetry.    We will see that such  integrated correlators,  introduced in  \cite{Binder:2019jwn}, can be explicitly determined by supersymmetric  localisation.  Not only do the large-$N$ expansions of these correlators make contact with the dual type IIB superstring amplitudes, but their properties at finite $N$ reproduce and generalise features determined directly from  Yang--Mills perturbation theory.

\subsection{Integrated correlators in $\mathcal{N}=4$ SYM}

We are interested in the correlation function of four superconformal primaries in the stress energy tensor supermultiplet, as well as `maximal $U(1)$-violating correlators'  of more than four operators. Explicitly, the four-point correlator in $\mathcal{N}=4$ SYM with arbitrary gauge group is given as
\begin{equation} \label{eq:4pts}
\!\!\!\!\! \langle \cO_2(x_1, Y_1)\dots \cO_2(x_4, Y_4)  \rangle = {1\over x_{12}^4 x_{34}^4} \left[\cT_{\rm free}(U,V;Y_i) + \mathcal{I}_4(U,V; Y_i) \cT(U,V) \right]  \, ,
\end{equation}
where the conformal invariant cross ratios $U, V$ are defined as 
\begin{equation}
U = \frac{x_{12}^2 x_{34}^2}{x_{13}^2 x_{24}^2}\, , \qquad V = \frac{x_{14}^2 x_{23}^2}{x_{13}^2 x_{24}^2}\, ,
\end{equation} 
and $\cO_2(x_i, Y_i)$ is the superconformal primary operator in the stress tensor supermultiplet of $\mathcal{N}=4$ SYM,  which is defined as
\begin{equation}
\cO_2(x_i, Y_i) = \tr ( \phi_I(x_i)  \phi_J(x_i) ) Y_i^I Y_i^J \, ,
\end{equation}
where $\phi_I$ is a scalar field in $\mathcal{N}=4$ SYM, and $\cO_2$ has conformal dimension $\Delta=2$. 
Here $I, J=1,2, \ldots, 6$ are the R-symmetry $SO(6)$ indices and $Y_i$ is a null polarization vector $Y_i \cdot Y_i=0$.  In writing down (\ref{eq:4pts}), we have used the partial non-renormalisation theorem of the correlator \cite{Eden:2000bk, Nirschl:2004pa}. This theorem implies that after separating out the free-theory contribution $\cT_{\rm free}$, the remaining part can be further factorised into a universal factor $\mathcal{I}_4(U,V; Y_i)$, which is fixed by the symmetries, and corresponds to  the supercharge $\delta^{16}(Q)$ in the superstring amplitude.  All the non-trivial dynamics of the correlator is contained in $\cT(U,V)$.  

Many properties of  $\cT(U,V)$ have been studied.  In perturbation theory, it was evaluated  at one and two loops in \cite{Gonzalez-Rey:1998wyj, Eden:1998hh, Eden:1999kh, Eden:2000mv, Bianchi:2000hn}, and at three loops in  \cite{Drummond:2013nda}.  The planar loop integrands have been constructed up to ten loops \cite{Bourjaily:2015bpz, Bourjaily:2016evz},  and  non-planar contributions first appear at  four loops.  The non-planar four-loop integrand was determined in \cite{Fleury:2019ydf}. These very high order results for the integrands were made possible by the discovery of a hidden permutation symmetry  \cite{Eden:2012tu, Eden:2011we}. In the strong coupling limit, the tree-level Witten diagrams for this correlator were computed in early days of AdS/CFT duality \cite{Maldacena:1997re, Gubser:1998bc, Witten:1998qj} \cite{DHoker:1999kzh, Liu:1998ty, DHoker:1999mqo, Arutyunov:2000py}, and the one-loop contribution in the supergravity limit was studied more recently \cite{Alday:2017xua, Aprile:2017bgs, Aprile:2017xsp, Alday:2017vkk}.\footnote{See  \cite{Heslop:2022qgf}  (chapter 8 of this SAGEX Review) for more details  of related topics. }

However, we are interested in the correlator at finite  complex Yang--Mills coupling $\tau=\theta/(2\pi) + {4\pi i /g_{_{YM}}^2}$, which is important  for making manifest $SL(2, \Z)$  self-duality of the theory and for the understanding its relation to the exact results of superstring amplitudes that were discussed in the previous section.  Although evaluating  a non-trivial correlation function at finite coupling is generally challenging,  powerful methods have recently been developed for determining integrated correlators based on supersymmetric localisation.   This utilises the fact that $\cN=4$ SYM can be expressed as the $m\to 0$ limit of $\cN=2^*$ SYM, a massive deformation of the $\cN=4$ theory where the hypermultiplet is given a mass $m$.   Using this fact the integral of the $\cN=4$ SYM   four-point correlator   (\ref{eq:4pts})  over the positions of the operators, $x_i$,  with a particular measure can be expressed in terms of properties of the partition function of $\cN=2^*$ SYM on $S^4$, which was evaluated some time ago using supersymmetric localisation  \cite{Pestun:2007rz}.   More explicitly, the following are two examples of  integrated correlators that are determined by the $m\to 0$ limit of  derivatives of the $\cN=2^*$ partition function $Z(\tau,\bar{\tau},m)$ \cite{Binder:2019jwn, Chester:2020dja}.  
\bea 
\label{eq:integrated1}
\!\!\!\!\!\!\!\!\!\!\!\!\!\!\!\!\!\!  \!\!\!\!\!\!  \!\!\!\!\!\! {1\over 4} \Delta_{\tau} \partial^2_m \log Z(\tau, \bar \tau, m) \Big{|}_{m=0} 
= -{8\over \pi} \int_0^{\infty} dr \int_0^{\pi} d\theta {r^3 \sin^2(\theta) \over U^2}  \cT(U,V) \, ,
\eea
and
\begin{align}
\label{eq:integrated2}
 \partial^4_m \log Z(\tau, &\bar \tau, m) \Big{|}_{m=0} \\
  &\notag= -{32\over \pi}  \int_0^{\infty} dr \int_0^{\pi} d\theta {r^3 \sin^2(\theta) \over U^2} (1+U+V) \bar{D}_{1111} (U, V)\cT(U,V)  +48\zeta(3) c\, ,
\end{align}
where $r^2=V$, $1- 2r \cos(\theta) +r^2 =U$, $\bar{D}_{1111}$ is the so-called $D$-function appearing in the computation of contact Witten diagrams (see e.g. the appendix D of \cite{Arutyunov:2002fh}), and $c$ denotes the central charge of the theory.  Roughly speaking, the four derivatives bring down four integrated operators, and setting $m=0$ at the end is necessary for reducing $\cN=2^*$ back to  $\cN=4$ SYM.

The localisation expression for $Z(\tau, \bar \tau, m)$ is given by a $N$-dimensional matrix model integral where the integrand  consists of the matrix model measure multiplying the product of two factors.  One factor is simply the one-loop determinant of the $\cN=2^*$ theory, which leads to the perturbative sector of the integrated correlators.  The other factor is the Nekrasov partition function that leads to the instanton contributions.

The $k$-instanton contribution can be expressed as a $k$-dimensional contour integral  \cite{Nekrasov:2002qd, Nekrasov:2003rj}.  Determining the explicit properties of  $Z(\tau, \bar \tau, m)$ is generally very complicated, especially at finite $\tau$ and for general  values of $N$.  However, the expressions for the integrated $\cN=4$ SYM correlators  \eqref{eq:integrated1} and  \eqref{eq:integrated2} depend only on the coefficients of the terms of order $m^2$ and order $m^4$  in the expansion of   $Z(\tau, \bar \tau, m)$ in powers of $m$.   This considerably simplifies some aspects of the analysis, but it is still very difficult to obtain the explicit results that display both the $\tau$ dependence and the $N$ dependence of the integrated correlators.  However, there are a number of results that address the large-$N$ expansion, both  with fixed 't Hooft coupling and with fixed $g_{_{YM}}$ \cite{Binder:2019jwn, Chester:2019jas, Chester:2020vyz, Chester:2020dja}. In the large-$N$ 't Hooft limit the instantons are exponentially  suppressed, which obscures the $SL(2,\Z)$ symmetry, but the large-$N$ expansion with fixed $g_{_{YM}}$ accounts for Yang--Mills instantons and makes $SL(2,\Z)$ explicit.  

Here we will review the arguments in  \cite{Dorigoni:2021bvj, Dorigoni:2021guq},  in which the first integrated correlator    \eqref{eq:integrated1}  is expressed as a two-dimensional lattice sum, which is an explicitly modular invariant function of  function of $\tau$ for all values of $N$.  We will also review the generalisation to $\cN=4$ SYM with an arbitrary classical gauge group    \cite{Dorigoni:2022zcr} (which made use of the analysis of the perturbative sector in \cite{Alday:2021vfb}).

We will also review results  for the second integrated correlator (\ref{eq:integrated2})  which are less complete since they only account for several low-lying terms in the large-$N$ expansion that were determined in \cite{Chester:2020vyz}.

The study of integrated correlators in $\mathcal{N}=4$ SYM has been further extended to maximal $U(1)_Y$-violating  (MUV)  integrated correlators \cite{Green:2020eyj, Dorigoni:2021rdo}, which are holographic duals of MUV amplitudes in type IIB superstring theory in an $AdS_5\times S^5$  background  that were discussed in the last section. 
 Here $U(1)_Y$ is the bonus  U(1) \cite{Intriligator:1998ig}, of  the gauge theory, which is  a true symmetry of the free theory but more generally is broken to a $\Z_4$ automorphism of the supergroup $PSU(2,2|4)$ and is dual to the (broken)  
  $U(1)$ R-symmetry of the type IIB superstring.
  
An example of  an unintegrated MUV correlator is 
\bea\label{eq:O2Otau}
\langle \cO_2(x_1, Y_1)\dots \cO_2(x_4, Y_4) \cO_{\tau}(x_5) \ldots \cO_{\tau}(x_{n})   \rangle \, , 
\eea 
where $\cO_{\tau}$ is the chiral Lagrangian of ${\cal N}=4$ SYM. Other types of MUV correlators are related to this one by supersymmetry and superconformal symmetry.  This transforms with  weight  $(n-4,4-n)$.\footnote{Note that the modular weights of MUV correlators have the opposite signs to the modular weights of the holographic dual  MUV amplitudes.}  The chiral Lagrangian $\cO_{\tau}$ is dual to the dilaton, and the holographic version of the soft-dilaton condition (\ref{eq:softstring}) takes the form of
\bea \label{eq:recursion} 
 &\cD_{n-5}  \langle \cO_2(x_1, Y_1)\dots \cO_2(x_4, Y_4)  \cO_{\tau}(x_5) \ldots   \cO_{\tau}(x_{n-1})  \rangle   \\
=&    {1\over 2}   \int d^4 x_{n} \langle \cO_2(x_1, Y_1)\dots \cO_2(x_4, Y_4) \cO_{\tau}(x_5) \ldots  \cO_{\tau}(x_{n-1}) \cO_{\tau}(x_{n})   \rangle \, . \nonumber
\eea 
When considering the perturbative contributions of the correlators this relation leads to a very efficient method for constructing loop integrands for the four-point correlator \cite{Dorigoni:2021rdo}, which  was utilised in \cite{Eden:2011we, Eden:2012tu, Bourjaily:2015bpz, Bourjaily:2016evz, Fleury:2019ydf} for calculating  the four-point correlator at high orders.   

  \setcounter{footnote}{0} 

\subsection{Exact results for integrated correlators}

We will now discuss  the exact structure of the integrated correlator (\ref{eq:integrated1})    that is proportional to $\Delta_{\tau} \partial^2_m \log Z_{G_N}(\tau, \bar \tau, m)|_{m=0}$.  We have introduced the subscript $G_N$ to the partition function to indicate that we are considering the gauge group to be any classical Lie group, so $G_N=SU(N)$ (as in  \cite{Dorigoni:2021bvj, Dorigoni:2021guq}) or   $SO(2N)$, $SO(2N+1)$,  $USp(2N)$ (as in \cite{Dorigoni:2022zcr}).  

Montonen--Olive duality\footnote{The term `Montonen--Olive' duality is often used interchangeably with `Goddard--Nuyts--Olive'  (GNO) duality \cite{Goddard:1976qe}.   GNO  demonstrated that in a Yang--Mills theory  that has magnetic monopoles and gauge group $G$, the magnetic charges are associated with points on the weight lattice of the dual group $^LG$. The superscript $^L$  indicates that the dual group is the Langlands group. Since here we consider only correlators of local operators, these depend only from the Lie algebra $\mathfrak{g}_N$  and its dual, $^L\mathfrak{g}_N$, and so global features of this duality are not relevant in this article.}
\cite{Montonen:1977sn, Witten:1978mh,  Osborn:1979tq} implies that when the gauge group is simply-laced ($SU(N)$ or $SO(2N)$) correlators must be invariant under $SL(2,\Z)$.  This is generated by the transformations $S:\, \tau\to -1/\tau$ and $T:\, \tau\to \tau+1$.  For the non simply-laced gauge groups, $SO(2N+1)$ and $USp(2N)$, duality is generated by the transformations $\hat S:\, \tau\to -1/(2\tau)$ and $T:\, \tau+1$.  The action of $\hat S$ transforms $SO(2N+1)$ into $USp(2N)$  and vice versa, so it is not a self-duality.   The transformations $\hat S \,T\, \hat S$ and $T$ generate the congruence subgroup  $\Gamma_0(2) \subset SL(2,\Z)$.  An element $\gamma \!=\! \begin{psmallmatrix}a & b\\c & d\end{psmallmatrix} \!\in\! SL(2,\mathbb{Z})$ belongs to $\Gamma_0(2)$ if $c = 0 \,{\rm mod}\ 2$.   So $\Gamma_0(2)$  is the self-duality group when the gauge groups are non simply-laced classical groups.

In \cite{Dorigoni:2022zcr}, it was found that the integrated correlators for any classical Lie group can be expressed in the compact and unified form,
\begin{align} 
\label{eq:mainres}
\mathcal{C}_{G_N}(\tau,\bar{\tau})  &\equiv {1\over 4} \Delta_{\tau} \partial^2_m \log Z_{G_N}(\tau, \bar \tau, m) \Big{|}_{m=0} \\
&=  \sum_{(m,n)\in \Z^2} \int_0^\infty dt\left(B^1_{G_N}(t) e^{-t \pi\frac{ |m+n\tau|^2}{\tau_2}}+B_{G_N}^2 (t) e^{-t\pi \frac{ |m+ 2 n \tau|^2}{ 2 \tau_2}}\right)\, ,\notag
\end{align}
where $B^1_{G_N}(t)$ and $B^2_{G_N}(t)$ are rational functions of $t$.   In the  simply-laced cases $B_{G_N}^2 (t) =0$, and  hence $\mathcal{C}_{G_N}(\tau,\bar\tau)$ is $SL(2, \Z)$ invariant. 
The rational function $B_{SU(N)}(t)$ was constructed in \cite{Dorigoni:2021bvj, Dorigoni:2021guq} and it is explicitly given by
\begin{equation}
\label{eq:BSUN}
B_{SU(N)}(t) = \frac{\mathcal{Q}_{SU(N)}(t)}{(t+1)^{2N+1}}\, ,
\end{equation}
and  $\mathcal{Q}_{SU(N)}(t)$ is a polynomial of degree $(2N-1)$ that  takes the form
 \begin{align}
&\mathcal{Q}_{SU(N)}(t)\label{polydef}
= -{1\over 4} N (N-1) (1-t)^{N-1} (1+t)^{N+1}  \\
 &\notag \left\{ \left(3+  (8N+3t-6) \, t\right ) P_N^{(1,-2)} \left(\frac{1+t^2}{1-t^2}\right)  + \frac{1}  {1+t}   \left(3t^2-8Nt-3 \right) P_N^{(1,-1)}    \left(\frac{1+t^2}{1-t^2}\right)  \right\} \,,
\end{align}
with $P_N^{(\alpha,\beta)} (z)$ being a Jacobi polynomial.   The function $B_{SU(N)}(t)$ satisfies several interesting relations
 \begin{align}
B_{SU(N)} (t) &= t^{-1}\,B_{SU(N)} (t^{-1})\, , \\
 \int_0^\infty B_{SU(N)}(t) dt &= {N(N-1) \over 8}\,,\qquad \int_0^\infty \frac{1}{\sqrt t} B_{SU(N)}(t) dt=0\,.
\label{eq:invert}
\end{align}
The first of these equations is an inversion relation that follows automatically from the lattice sum definition of the integrated correlator \eqref{eq:mainres}, as was pointed out in \cite{Collier:2022emf} (where the lattice sum was re-expressed in terms of a modular invariant spectral represnetation).

For the non simply-laced cases, GNO duality interchanges the $SO(2N+1)$ theory with the  $USp(2N)$ theory. 
This is a property of the expression (\ref{eq:mainres}) by virtue of the fact that for these groups the coefficient functions satisfy the relations 
\bea \label{eq:SOSp}
B_{SO(2N+1)}^1 (t) = B_{USp(2N)}^2 (t)\, , \qquad B_{SO(2N+1)}^2 (t) = B_{USp(2N)}^1 (t) \,,
\eea 
which ensure that  the action of  $\hat S$ interchanges the two terms, and therefore relates $\mathcal{C}_{SO(2N+1)}(\tau,\bar\tau)$ with $\mathcal{C}_{USp(2N)}(\tau,\bar\tau)$. 

It is also notable that $B_{SU(-N)}(t) = B_{SU(N)}(-t)$ which is directly connected to the relation $\mathcal{C}_{SU(N)}(\tau,\bar{\tau}) = \mathcal{C}_{SU(-N)}(-\tau,-\bar{\tau})$.
Exact expressions for all $B^1_{G_N}(t), B^2_{G_N}(t)$ can be found in \cite{Dorigoni:2022zcr}, however, as we will shortly argue,  $\cC_{SO(n)}$ and $\cC_{USp(2N)}$ can be expressible in terms of rational linear combinations of $\cC_{SU(m)}$.  Furthermore, we will also see that $\cC_{SU(m)}$ can be expressed in terms of $\cC_{SU(2)}$.

Using the definition of the non-holomorphic  Eisenstein series (\ref{eq:lattice})  the integrated correlator can be written as the formal expansion
\bea
\label{eq:cgeis}
\mathcal{C}_{G_N}(\tau,\bar{\tau}) = -b_{G_N}(0) + \sum_{s=2}^\infty  \left[ b_{G_N}^1 (s)\, E(s;\tau,\bar\tau)  + b_{G_N}^2 (s)\,E(s;2\tau,2\bar\tau)\right] \,  ,
\label{eisenform}
\eea
where the coefficients $b^1_{G_N}(s)$ and $b^2_{G_N}(s)$ are rational numbers that are determined by the power series expansion of  $B^i_{G_N}(t)$ in the form
\bea
B_{G_N}^i (t) = \sum_{s=2}^\infty \frac{b_{G_N}^i (s)}{\Gamma(s)}\,\, t^{s-1}\,, \qquad \qquad i=1,2 \, ,
\label{Bexpand}
\eea
and we defined $b_{G_N}(0) = b_{G_N}^1(0)+b_{G_N}^2(0)$.

Again for the simply-laced cases $b_{G_N}^2 (s)=0$, and for the non simply-laced cases
\bea
b_{SO(2N+1)}^1 (s)  = b_{USp(2N)}^2 (s)  \,  , \qquad  b_{SO(2N+1)}^2 (s)  = b_{USp(2N)}^1 (s) \, , 
\eea
which manifest GNO duality. 
In \cite{Collier:2022emf}  the formal expansion (\ref{eq:cgeis})  was expressed in terms of a spectral decomposition for  $\mathcal{C}_{SU(N)}$, and a  similar expression was given for general $\cC_{G_N} $ in  \cite{Dorigoni:2022zcr}.

Remarkably, the integrated correlators satisfy Laplace-difference equations that relate correlators of different gauge groups. In the case of $SU(N)$, this takes the form
\bea \label{eq:lapdiffSU}
\!\!\! \!\!\! \!\!\! \!\!\! \!\!\!  \Delta_\tau \mathcal{C}_{SU(N)}(\tau,\bar{\tau})    -4c_{SU(N)} & \left[ \mathcal{C}_{SU(N+1)}(\tau,\bar{\tau}) - 2 \mathcal{C}_{SU(N)}(\tau,\bar{\tau}) + \mathcal{C}_{SU(N-1)}(\tau,\bar{\tau})  \right] \cr
  &\!\!\!\!\!\!\!\!\!\!\!\!\!\!\!\!\!\!\!\!\!-(N+1) \mathcal{C}_{SU(N-1)} (\tau,\bar{\tau}) + (N-1) \mathcal{C}_{SU(N+1)}(\tau,\bar{\tau}) =0 \, ,
\eea
where $c_{SU(N)} = (N^2-1)/4$ is the central charge.  This is a powerful equation that determines $\cC_{SU(N)}(\tau,\bar{\tau})$ for general values of $N$ in terms of $\cC_{SU(2)}(\tau,\bar{\tau})$, once the boundary condition $\cC_{SU(1)}(\tau,\bar{\tau})=0$ is imposed.

The equations for other gauge groups take similar forms, Thus, the Laplace-difference equation for the $SO(n)$ correlator  (where $n=2N$ or $2N+1$)  is given by
\bea
\!\!\!\! \!\!\!\! \!\!\!\! \!\!\!\!  \Delta_\tau \mathcal{C}_{SO(n)}(\tau,\bar{\tau})   -  2 c_{SO(n)} &  \left[ \mathcal{C}_{SO(n+2)}(\tau,\bar{\tau}) -2 \mathcal{C}_{SO(n)}(\tau,\bar{\tau}) +\mathcal{C}_{SO(n-2)} (\tau,\bar{\tau})  \right] \cr
&\!\!\!\!\!\!\!\!\!\!\!\! - n\, \mathcal{C}_{SU(n-1)} (\tau,\bar{\tau}) +(n-1) \mathcal{C}_{SU(n)} (\tau,\bar{\tau}) =0 \, ,
\label{eq:lapdiffSO}
\eea
where  $c_{SO(n)} = n(n-1)/8$ is the central charge for $SO(n)$.   This equation relates  $SO(n)$ and $SU(n)$ correlators.
The $SO(3)$ case is an exception since the Dynkin index of $SO(n)$ is discontinuous as $n=3$, hence in that case the integrated correlator is {actually given by  $\mathcal{C}_{SO(3)}(\frac{\tau}{2},\frac{\bar\tau}{2})$ (rather than $\mathcal{C}_{SO(3)}(\tau,\bar\tau)$), which agrees with the result of supersymmetric localisation \cite{Alday:2021vfb}.
For $USp(n)$ correlators (where $n=2N$), we have
\bea
\!\!\!\! \!\!\!\! \!\!\!\! \!\!\!\!  \!\!\!\! \!\!\!\!  \Delta_\tau \mathcal{C}_{USp(n)}(\tau,\bar{\tau}) - 2 c_{USp(n)} & \left[ \mathcal{C}_{USp(n-2)}(\tau,\bar{\tau}) -2 \mathcal{C}_{USp(n)}(\tau,\bar\tau)+\mathcal{C}_{USp(n+2)} (\tau,\bar{\tau}) \right] \cr
&  \!\!\!\!\!\!\!\!\!\!\!\!+   n\, \mathcal{C}_{SU(n+1)} (2\tau,2\bar{\tau})- (n+1)\mathcal{C}_{SU(n)} (2\tau,2\bar{\tau})   =0\, ,
\label{eq:lapdiffUSp}
\eea
where  $c_{USp(n)} = n(n+1)/8$ is the central charge for $USp(n)$. This equation relates $USp(n)$  and $SU(n)$ correlators.    The localisation expression for the correlator can be used to show that 
\bea  \label{eq:isometry}
\mathcal{C}_{SO(3)}\Big(\frac{\tau}{2},\frac{\bar\tau}{2}\Big) = \mathcal{C}_{USp(2)}(\tau,\bar{\tau}) = \mathcal{C}_{SU(2)}(\tau,\bar{\tau}) \, ,
\eea
which is consistent with the isomorphism of the corresponding Lie algebras.
Combining this initial condition with the fact that $\cC_{SU(N)}(\tau,\bar\tau)$ is determined by (\ref{eq:lapdiffSU}), it is then straightforward to show that the Laplace-difference equations (\ref{eq:lapdiffSO}) and (\ref{eq:lapdiffUSp}) determine $\mathcal{C}_{SO(n)}$ and $\mathcal{C}_{USp(n)}$ in terms of finite rational linear combinations of $\mathcal{C}_{SU(m)}$ correlators.  For example,
\begin{align}
&\mathcal{C}_{SO(4)}(\tau,\bar\tau) = 2\, \mathcal{C}_{SU(2)}(\tau,\bar\tau) \, , \qquad \qquad \mathcal{C}_{SO(6)}(\tau,\bar\tau) =  \mathcal{C}_{SU(4)}(\tau,\bar\tau) \, , \\
&\notag\mathcal{C}_{SO(8)}(\tau,\bar\tau) = \\
&\notag{-}2\,\mathcal{C}_{SU(2)}(\tau,\bar\tau )+\frac{8}{3} \mathcal{C}_{SU(3)}(\tau,\bar\tau )-2 \,\mathcal{C}_{SU(4)}(\tau,\bar\tau )+\frac{4}{5} \mathcal{C}_{SU(5)}(\tau,\bar\tau)+\frac{2}{3} \mathcal{C}_{SU(6)}(\tau,\bar\tau)\, .
\end{align}
Since $\mathcal{C}_{SU(m)}(\tau,\bar{\tau})$ is invariant under $SL(2,\mathbb{Z})$ for all $m\in\mathbb{N}$, it follows that $\mathcal{C}_{SO(2N)}(\tau,\bar{\tau})$ is also invariant under $SL(2,\mathbb{Z})$, as expected from GNO duality for $SO(2N)$.

Similarly, 
\begin{align}
\mathcal{C}_{SO(5)}(\tau,\bar\tau)= \left[ -2\, \mathcal{C}_{SU(2)}(\tau,\bar\tau )  + \frac{4}{3}\mathcal{C}_{SU(3)}(\tau,\bar\tau ) \right] +  \left[-2\, \mathcal{C}_{SU(2)}(2 \tau,2\bar\tau
   ) +\frac{4}{3} \mathcal{C}_{SU(3)}(2\tau,2\bar\tau ) \right] \, ,
   \label{so5cor}
\end{align}
with an identical result for $\mathcal{C}_{USp(4)}(\tau,\bar\tau)$, reflecting the fact that $USp(4) \cong SO(5)$.  It is instructive to compare  $\mathcal{C}_{SO(7)}$ and $\mathcal{C}_{USp(6)}$, in order to get some insight into the  way in which  
GNO duality that relates $\mathcal{C}_{SO(2N+1)}$ and $\mathcal{C}_{USp(2N)}$ is realised.  From  \eqref{eq:lapdiffSO} we find 
\begin{align}
\mathcal{C}_{SO(7)}(\tau,\bar\tau) &\notag= \left[\frac{8}{5} \mathcal{C}_{SU(2)}(\tau,\bar\tau )-\frac{12}{5} \mathcal{C}_{SU(3)}(\tau,\bar\tau
   )+\frac{3}{5} \mathcal{C}_{SU(4)}(\tau,\bar\tau )+\frac{4}{5} \mathcal{C}_{SU(5)}(\tau,\bar\tau ) \right] \\
   &   \label{so7cor} + \left[ \frac{3}{5}\mathcal{C}_{SU(2)}(2 \tau,2\bar\tau )-\frac{12}{5} \mathcal{C}_{SU(3)}(2 \tau,2\bar\tau )+\frac{8}{5} \mathcal{C}_{SU(4)}(2 \tau,2\bar\tau) \right]\, ,
\end{align}
and from \eqref{eq:lapdiffUSp},
\begin{align}
\mathcal{C}_{USp(6)}(\tau,\bar\tau) &\notag= \left[\frac{8}{5} \mathcal{C}_{SU(2)}(2\tau,2 \bar\tau )-\frac{12}{5} \mathcal{C}_{SU(3)}(2\tau,2\bar\tau
   )+\frac{3}{5} \mathcal{C}_{SU(4)}(2\tau, 2\bar\tau )+\frac{4}{5} \mathcal{C}_{SU(5)}(2\tau,2\bar\tau ) \right] \\
   & + \left[ \frac{3}{5}\mathcal{C}_{SU(2)}(\tau,\bar\tau )-\frac{12}{5} \mathcal{C}_{SU(3)}(\tau,\bar\tau )+\frac{8}{5} \mathcal{C}_{SU(4)}(\tau,\bar\tau) \right]\, .
   \label{USp6cor}
\end{align}
Since $ \mathcal{C}_{SU(N)}(\tau,\bar\tau)= \mathcal{C}_{SU(N)}(-\frac{1}{\tau},-\frac{1}{\bar\tau})$ and $\mathcal{C}_{SU(N)}(2\tau, 2\bar\tau) =\mathcal{C}_{SU(N)}(-\frac{1}{2\tau},-\frac{1}{2\bar\tau})$, it follows from \eqref{so7cor} and \eqref{USp6cor} that under the transformation $\hat S: \tau \to   - 1/(2\tau) $, $\mathcal{C}_{SO(7)}(\tau,\bar\tau)$ transforms into  $\mathcal{C}_{USp(6)}(\tau,\bar\tau)$. More generally, by induction, using the Laplace-difference equations \eqref{eq:lapdiffSO} and \eqref{eq:lapdiffUSp},  one can prove 
\begin{equation} \label{eq:GNO}
\mathcal{C}_{SO(2N+1)}(\tau, \bar{\tau}) = \mathcal{C}_{USp(2N)}\Big(-\frac{1}{2\tau},-\frac{1}{2\bar\tau}\Big) \, ,
\end{equation}
which is the statement of GNO duality, recalling our previous comment that for $N=1$ the localised correlator equals $\mathcal{C}_{SO(3)}(\frac{\tau}{2},\frac{\bar{\tau}}{2})$, which also coincides with the integrated correlators $\mathcal{C}_{SU(2)}(\tau,\bar{\tau}) = \mathcal{C}_{USp(2)} (\tau,\bar{\tau})$.

 \setcounter{footnote}{0} 

\subsection{SYM perturbation theory}

Starting from our basic expression \eqref{eq:mainres} for $\cC_{G_N}(\tau,\bar{\tau})$  it is straightforward to evaluate the  perturbation expansion $\cC_{G_N}^{pert}(\tau_2)$ in the region  $\tau_2 = 4\pi / g_{YM}^2 \to \infty$, which agrees with the localisation result originally derived in   \cite{Alday:2021vfb}.   This perturbative expansion
can be organised in a striking manner  by defining suitable  expansion parameters, $a_{G_N}$,  for each gauge group.  These generalisations of  the 't Hooft coupling are given by 
\bea  
\!\!\!\!\!\!\!\!\!\!\!\! a_{SU(N)} = {N g_{_{YM}}^2 \over 4\pi^2}\,, \qquad
a_{SO(n)} ={(n-2)  g_{_{YM}}^2 \over 4\pi^2} \,, \qquad
a_{USp(n)} = {(n+2) g_{_{YM}}^2  \over 8\pi^2} \, ,
\label{pertgroup}
\eea
where $n=2N$ or $2N+1$ for $SO(n)$, and $n=2N$ for $USp(n)$. The $SU(N)$ coupling is  the standard 't Hooft parameter  (rescaled by  $4\pi^2$), while $a_{SO(n)}$ and $a_{USp(n)}$ are the generalisations for $SO(n)$ and $USp(n)$ theory (see also \cite{Cvitanovic:2008zz}). Note the parameters $a_{G}$ defined in \eqref{pertgroup} can be rewritten in a unified form  $a_{G} = \frac{ {h^\lor_G} g_{_{YM}}^2}{4\pi^2} $, with $ {h^\lor_G} $ the dual Coxeter number for the group $G$. The appearance of the dual Coxeter number is  natural since in $\mathcal{N}=4$ SYM all fields belong to the adjoint representation. \footnote{As already mentioned the case of $SO(3)$ is special and one needs to rescale $g_{_{YM}} \to \sqrt{2}\, g_{_{YM}}$ and define $a_{SO(3)} = {g_{_{YM}}^2 / (2\pi^2)}$ so that $a_{SO(3)} = a_{SU(2)} = a_{USp(2)}$.  We furthermore have $a_{SU(4)}=a_{SO(6)}$ and $a_{USp(4)}=a_{SO(5)}$, consistent with the isomorphic relations among the corresponding algebras.  }

In terms of these parameters we find that the perturbative expansion of all the integrated correlators can be expressed in the following form, 
\begin{align}
\cC_{G_N}^{pert} (\tau_2)&\notag =   \, -4 c_{G_N} \left[ \frac{3   \, \zeta (3) a_{G_N}   }{2} -\frac{75 \, \zeta (5)a_{G_N}^2}{8} 
+\frac{735 \,\zeta (7) a_{G_N}^3}{16} -\frac{6615  \,\zeta (9)  \left(1 + P_{G_N, 1}\right)  a_{G_N}^4 }  {32} \right. \\
& \label{pertexp} \left. +\, \frac{114345 \,  \zeta (11) \left(1+  P_{G_N, 2}  \right)a_{G_N}^5  }{128 } + O(a_{G_N}^{6}) \right] \, .
\end{align} 
A striking feature is that the first three perturbative contributions are universal and their dependence on $N$ is contained entirely within the central charge $c_{G_N}$ and $a_{G_N}$. Explicit ``non-planar''  factors, $P_{G_N,i}$, where $i = \ell-3$ and $\ell$ is the loop number, first  enter at four loops and the first few examples are listed below:
\begin{align}
\label{suparam}
\qquad P_{SU(N), 1}  &= \frac{2}{7N^2} \, , \,\,\,\qquad\qquad\qquad\qquad P_{SU(N), 2} =  {1 \over N^{2}} \, , \cr
P_{SO(n), 1}  &= -\frac{n^2-14 n+32}{14 (n-2)^3}\, , \qquad\qquad P_{SO(n), 2} = -\frac{n^2-14 n+32}{8 (n-2)^3}\, \,, \cr
P_{USp(n), 1}  &= \frac{n^2+14 n+32}{14 (n+2)^3}\ \, ,\,\qquad \qquad P_{USp(n), 2} = \frac{n^2+14 n+32}{8 (n+2)^3} \, ,
\end{align}
and further details and higher-order terms are given in  \cite{Dorigoni:2022zcr}.

From \eqref{suparam} we see that for $SU(N)$  (\ref{suparam}) is the well-known genus-expansion in powers of $1/N^2$ and $a_{SU(N)}$  \cite{tHooft:1973alw}. However there seems to be no systematic analysis of the analogous expansions for $SO(n)$  and $USp(n)$ (see \cite{Cvitanovic:2008zz} for some limited results). Given the expressions in \eqref{suparam}, as well as higher orders presented in \cite{Dorigoni:2022zcr}, we see that the large-$N$ expansions for $SO(n)$ (with $n=2N$ or $n=2N+1$) and $USp(n)$ (with $n=2N$) are  expressed purely in powers of $1/(n-2)$ and $1/(n+2)$, respectively. 

From \eqref{pertexp}  we see that the planar contribution is the same for all gauge groups, while the non-planar contributions only enter at $\ell\geq4$ loops. This property is consistent  with the construction of perturbative loop integrands using the methods in \cite{Eden:2011we, Eden:2012tu}, and provides important information concerning  large-$N$ expansions.

We note that the precise coefficients of the perturbative expansion \eqref{pertexp} can be verified using standard Feynman diagram computations. In particular the first two loops were computed in \cite{Dorigoni:2021guq} while the planar terms up to order $O(a_{G_N}^4)$ were derived in \cite{Wen:2022oky} by understanding that the Feynman integrals associated with the integrated correlator are simply periods of certain conformal Feynman graphs, for which  special calculational techniques are available.  These results make use of the perturbative loop integrands constructed in \cite{Eden:2011we, Eden:2012tu, Bourjaily:2015bpz, Bourjaily:2016evz} and the precise expression for the integrated correlator \eqref{eq:integrated1}. 
 
The perturbative expansion \eqref{pertexp} and the non-planar expressions \eqref{suparam} are consistent with certain symmetries.
In particular, for $SU(N)$ we have 
\begin{equation}
c_{SU(N)} = c_{SU(-N)} \, , \qquad a_{SU(N)}  = a_{SU(-N)}\, , \qquad  P_{SU(N), i} = P_{SU(-N), i} \, ,
\end{equation}
hence
\begin{equation}
\label{eq:suct}
\cC_{SU(N)}^{pert}(g^2_{_{YM}}) =\cC_{SU(-N)}^{pert}(-g^2_{_{YM}})\,.
\end{equation}
 Similarly for $SO(2N)$ and $USp(2N)$ we notice
\begin{equation}
  c_{SO(2N)} = c_{USp(-2N)} \, , \qquad a_{SO(2N)}  = 2 a_{USp(-2N)} \,, \qquad P_{SO(2N), i} = P_{USp(-2N), i} \,,   
\end{equation}
which lead to 
\begin{equation}
\label{eq:soct}
\cC_{SO(2N)}^{pert}( g^2_{_{YM}}) =\cC_{USp(-2N)}^{pert}(-2 g^2_{_{YM}}) \, .
\end{equation}
These relations have been further checked at higher orders and are furthermore consistent with the Laplace-difference equations \eqref{eq:lapdiffSU}, \eqref{eq:lapdiffSO}, and \eqref{eq:lapdiffUSp}.  

The relations (\ref{eq:suct}) and (\ref{eq:soct})  also hold for the perturbative expansion of the other localised integrated correlator  that was  defined in (\ref{eq:integrated2}) and is proportional to  $\partial^4_m \log Z_{G_N}(m, \tau, \bar \tau) \big{|}_{m=0}$. And the perturbative terms of this integrated correlator have also been verified to match the explicit Feynman diagram calculations  \cite{Wen:2022oky}.

\subsection{Maximal $U(1)_Y$-violating correlators}
Given the exact results for the integrated four-point correlators \eqref{eq:mainres},   the $n$-point maximal $U(1)_Y$-violating correlators defined in \eqref{eq:O2Otau}  can be obtained for any classical gauge group  by acting  on $\cC_{G_N}(\tau,\bar\tau)$  with modular covariant derivatives defined in \eqref{covderdef}.

An integrated version of the MUV correlators, can be obtained starting from the integrated correlator \eqref{eq:integrated1}
$\mathcal{C}_{G_N}(\tau,\bar\tau) \equiv \mathcal{C}^{(0)}_{G_N}(\tau,\bar\tau)$ and inserting multiple factors of the integrated chiral Lagrangian, $\int \!dx\, \mathcal{O}_\tau(x)$.  Such insertions are obtained by applying multiple covariant derivatives $\mathcal{D}_w$ to $\mathcal{C}_{G_N}(\tau,\bar\tau)$. The resulting expression is a $(w,-w)$ modular form given by
 \begin{equation}
\mathcal{C} ^{(w)}_{G_N}(\tau,\bar\tau) = 2^w\, \mathcal{D}_{w-1} \mathcal{D}_{w-2} \cdots \mathcal{D}_{0}\, \mathcal{C}^{(0)}_{G_N} (\tau,\bar\tau)  \, .
\label{muvres}
\end{equation}

Given that the Laplacian operators $\Delta_{(\mp)w}$, defined in \eqref{eq:Lap}  are Casimir operators on the vector space of modular forms $M_{w,-w}$, they commute with the covariant derivatives $\mathcal{D}_w$. Furthermore, since these Laplacians reduce to the standard one $\Delta_{0} = \Delta_{\tau}$,  on the space of modular invariant functions $M_{0,0}$  we can use \eqref{eq:lapdiffSU} to derive a system of Laplace-difference equations satisfied by maximally $U(1)_Y$-violating integrated correlators.
With the help of the explicit forms of $\Delta_{(\mp)w}$ in \eqref{eq:Lap}, we obtain the two equivalent Laplace-difference equations. To illustrate the idea, we will focus on the $SU(N)$ case in the following discussion. In particular, we find the $SU(N)$ MUV integrated correlators obey the following Laplace-difference equations, 
\begin{align} \label{eq:lapdiffSUw1}
  &\Big(4 \mathcal{D}_{w-1}\bar{\mathcal{D}}_{-w} +w(w-1)\Big)   \mathcal{C}_{SU(N)}^{(w)} -4c_{SU(N)}  \left(  \mathcal{C}_{SU(N+1)}^{(w)}   - 2   \mathcal{C}_{SU(N)}^{(w)}  +  \mathcal{C}_{SU(N-1)}^{(w)}   \right)\cr
  &-(N+1)  \mathcal{C}_{SU(N-1)}^{(w)} + (N-1)  \mathcal{C}_{SU(N+1)}^{(w)}  =0 \, ,
\end{align}
and
\begin{align} \label{eq:lapdiffSUw2}
& \Big(4 \bar{\mathcal{D}}_{-w-1}\mathcal{D}_{w} +w(w+1) \Big)   \mathcal{C}_{SU(N)}^{(w)}  -4c_{SU(N)}  \left(  \mathcal{C}_{SU(N+1)}^{(w)}  - 2   \mathcal{C}_{SU(N)}^{(w)}+  \mathcal{C}_{SU(N-1)}^{(w)}   \right)\cr
  &-(N+1)  \mathcal{C}_{SU(N-1)}^{(w)}+ (N-1)  \mathcal{C}_{SU(N+1)}^{(w)}   =0 \, .
\end{align}

The structure of the instanton and anti-instanton contributions is of particular interest \cite{Dorigoni:2021rdo}. Starting from the exact expression \eqref{eq:mainres} for $\mathcal{C}_{SU(N)}^{(0)}$ it is fairly straightforward to obtain the $k$-instanton and $k$-anti-instanton sectors of $\mathcal{C}_{SU(N)}^{(w)}$. Since for $w>0$ we know that $\mathcal{C}_{SU(N)}^{(w)}$ has modular weight $(w,-w)$, it follows that the $k$-instanton and  $k$-anti-instanton  contributions do not coincide (whereas they do in the $w=0$ case).

The precise results obtained for various values of $N$ in \cite{Dorigoni:2021rdo} are in accord with expectations from the analysis of semi-classical instanton contributions to MUV correlators in special cases treated in, for example, \cite{Green:1997me, Dorey:1999pd, Green:2002vf},   These references are all restricted  to leading orders in the $1/N$ expansion of $\mathcal{N}=4$ SYM correlators or  to the holographically related terms in the low-energy expansion of superstring amplitudes.   However, the present results go far beyond the semi-classical approximation and apply to any value of $N\ge 2$, but nevertheless some general features are explained by the leading order calculations.  

For example, the fact that the leading power of $g_{_{YM}}^2 \sim \tau_2^{-1}$ in the instanton background is of order $\tau_2^{w}$ is a direct reflection of the presence of $16$ superconformal zero modes.   The counting of powers of $\tau_2$ to leading order in $1/\tau_2$ is as follows.  The instanton profile of each operator insertion  involves the product of ($2\Delta-4w$) fermionic zero modes (where $\Delta$ is the dimension of the operator), each contributing a power $\tau_2^{-1/4}$, in addition to the power of $\tau_2$ in the normalisation of each operator.  The leading order instanton contribution to the $n$-point correlator necessarily  absorbs all 16 superconformal  fermion zero modes and is therefore of order $\tau_2^{n-16\times 1/4} = \tau_2^w$ as $\tau_2\to \infty$.  More explicitly, the instanton profile of the operator $\mathcal{O}_2(x)$ ($\Delta=2$, $w=0$) has four fermionic zero modes, while $\mathcal{O}_\tau(x)$ ($\Delta=4$, $w=2$) has no fermionic zero modes, and so the $k$-instanton sector for $\mathcal{C}_{SU(N)}^{(w)}$ behaves as
\begin{equation}
\langle  \mathcal{O}_2(x_1, Y_1) \cdots  \mathcal{O}_2(x_4, Y_4)\, \mathcal{O}_\tau(x_5)\cdots \mathcal{O}_\tau(x_{w+4})\rangle  \sim e^{ 2\pi i k \tau} \tau_2^w \, .
\label{ourcorr}
\end{equation}

A similar, albeit slightly more involved argument, allows us to deduce that instead the $k$-anti-instanton sector for $\mathcal{C}_{SU(N)}^{(w)}$ behaves as
\begin{equation}\label{eq:AntiIn}
\langle  \mathcal{O}_2(x_1, Y_1) \cdots  \mathcal{O}_2(x_4, Y_4)\, \mathcal{O}_\tau(x_5)\cdots \mathcal{O}_\tau(x_{w+4})\rangle   \sim e^{ -2\pi i k \bar{\tau}}\tau_2^{-w} \,.
\end{equation}
Thanks to our exact formula \eqref{eq:mainres} specialised to \eqref{muvres}, both of these statements can be verified for general values of $N$.

\subsection{$\cN =4$ SYM correlators at large-$N$ and  superstring amplitudes}
\label{sec:stringamps}

The large-$N$ expansion of $\cN=4$ $SU(N)$ SYM correlators makes contact with type IIB superstring theory in an $AdS_5\times S^5$ background. From the string theory perspective  this background is identified with the near-horizon limit of $N$  coincident $D3$-branes in the large-$N$ limit.  The Yang--Mills parameters are identified with the type IIB superstring parameters by the relations
\bea
\label{eq:dictionary}
g_s= \frac{g_{_{YM}}^2}{4\pi}\,, \qquad\qquad \frac{(\alpha')^2} {L^4} = \frac{1}{g_{_{YM}}^2 N}\, .
\eea

For  $\cN=4$ SYM with $G_N$ gauge group, where $G_N$ is $SO(2N)$, $SO(2N+1)$  or $USp(2N)$ the large-$N$ theory is holographically dual to type IIB superstring theory in  $AdS_5\times RP^5$.  This background is the near-horizon geometry of  an orientifold of $N$ coincident $D3$-branes in the large-$N$ limit  that are also coincident with an $O3$-plane, In this background string world-sheets  are non-orientable.  There are 
various types of $O3$-plane that carry different amounts of Neveu--Schwarz Neveu--Schwarz and Ramond-Ramond flux.  Depending on the choice of $O3$-brane the dual gauge theory has gauge group $SO(2N)$, $SO(2N+1)$ or $USp(2N)$.
The RR five-form flux associated with the various backgrounds is  
\begin{equation}\label{eq:Fluxes}
 \fluxSU   :=  N\,,\qquad\qquad \,\fluxSO   :=  \frac{n}{2}-\frac{1}{4}\,,\qquad\qquad\,\fluxUSp := \frac{n}{2}+\frac{1}{4}\,,
 \end{equation} 
 where $n=2N$ or $2N+1$ for $SO(n)$ and $n=2N$ for $USp(n)$.    

Two versions of the large-$N$  limit are of interest: 

(a) The 't Hooft limit in which $\lambda_{G_n} = g_{_{YM}}^2\, \flux$ is fixed.  In this limit Yang--Mills instantons are suppressed exponentially in $N$.  The $1/N$ expansion  generalises the conventional 't Hooft genus expansion and corresponds to the string perturbation expansion in the holographically dual string theory. This expansion is presented in detail in \cite{Dorigoni:2022zcr}, and can be used to obtain non-perturbative string corrections similar to \cite{Aniceto:2015rua,Dorigoni:2015dha,Arutyunov:2016etw}, but we will not review it here.

(b)  The large-$N$ fixed-$\tau$ expansion in which  Yang--Mills instantons are not suppressed and play a crucial r\*ole in making S-duality manifest.  In this case the expansion in powers of $1/N$ correspond to the low-energy expansion in the dual type IIB string theory. In other words, the leading term, which is of order $N^2$  is the dual of the supergravity term, the next term is of order $N^{\half}$ and is the dual of the $R^4$ term and so on.  We will now consider the  holographic interpretation of this  expansion of the integrated correlators in more detail.

Integrated correlators carry much less detailed information than unintegrated correlators since they have no space-time dependence.  However, the constraints of maximal supersymmetry are so strong  that  one can reconstruct some aspects of  the large-$N$ expansion of an unintegrated correlator (with its spacetime dependence) from knowledge of the large-$N$  expansions of  integrated correlators.  As a result, it is possible to check the holographic correspondence with the low-energy expansion of type IIB superstring amplitudes for the first few orders in the large-$N$ expansion as will now be explained.

Focusing for brevity just on  large-$N$ expansion of the $SU(N)$ theory, the unintegrated four-point correlator has the following simple analytic structure in Mellin space.  The Mellin amplitude $\mathcal{M}(s, t)$ is defined as \cite{Mack:2009mi, Penedones:2010ue}, 
\begin{equation} \label{eq:Mellin}
    \cT(U,V) = \int^{i \infty}_{-i \infty} {ds dt \over (4\pi i)^2} U^{s \over 2} V^{{u\over 2}-2} \Gamma\left[2 -{s\over 2} \right]^2
 \Gamma\left[2 -{t\over 2} \right]^2  \Gamma\left[2 -{u\over 2} \right]^2 \mathcal{M}(s, t) \, ,
\end{equation}
where $u=4-s-t$.\footnote{The Mellin variables $s,t,u$ should not be confused with the Mandelstam variables in the previous section. In the flat-space limit \cite{Penedones:2010ue}, in which $s, t, u \rightarrow \infty$, they do become the Mandelstam variables of scattering amplitudes after a suitable rescaling. } The large-$N$ expansion or large-central charge expansion of the Mellin amplitude takes the following simple form,\footnote{Here we have simply written down the most general expression with permutation symmetry at each order according to its power counting from its holographic dual. For instance, the $N^{-\frac{1}{2}}$ term is dual to $d^4R^4$ in $AdS_5 \times S^5$, which has four derivatives (note we have removed $R^4$ part by factoring out $\mathcal{I}_4(U, V; Y_i)$ in \eqref{eq:4pts}). The corresponding Mellin amplitude is then given by linear combination of $s^2+t^2+u^2$ and a constant \cite{Penedones:2010ue}. } 
\begin{align}
\label{eq:4ptsMst}
&\mathcal{M}(s, t)=  \\
& N^2 { \tilde{f}_0  \over (s-2)(t-2)(u-2) } + N^{{1\over 2} }f_1 + {\cal M}^{SG}(s,t) +  N^{-{1\over 2} } \left[ f_{2,1} (s^2+t^2+u^2) +  f_{2,2} \right]  \notag \\
& +N^{-{1} } \left[  f_{3,1}s tu+  f_{3,2} (s^2+t^2+u^2) +  f_{3,3} \right] + O(N^{-{3\over 2} } ) \notag\, .
\end{align}
The leading term is proportional to $N^2$ (or alternatively $c$ the central charge) and corresponds to the tree-level supergravity contribution in $AdS_5 \times S^5$, and ${\cal M}^{SG}(s,t)$ is the one-loop supergravity.  These terms are independent of $\tau$. We are interested in terms proportional to $N^{{1\over 2} }$, $N^{-{1\over 2} } $ and $N^{-{1} }$ that are stringy higher-derivative corrections proportional to  $R^4$, $d^4R^4$ and $d^6R^4$, respectively.  The coefficients of these string corrections $ f_1$, $f_{2,i}$ and $f_{3,i}$ are non-trivial (non-holomorphic) functions of the coupling $\tau$, and, as we will show later, these may be fixed using the large-$N$ expansion of the integrated correlators will discussed in the following.

\subsubsection*{The large-$N$ expansion of  $\cC_{SU(N)}$ at fixed $\tau$.}

The first few terms in the large-$N$ expansion of  the first  integrated correlator  (\ref{eq:integrated1}) with fixed $\tau$ was first studied for the $SU(N)$ case  in \cite{Chester:2019jas}   and then to any prescribed  order in $1/N$  by  making use of the exact expression or by solving the Laplace-difference equation \eqref{eq:lapdiffSU} \cite{Dorigoni:2021guq}, giving 
 \begin{align} 
 \label{oldexpand}
& \cC_{SU(N)} (\tau, \bar \tau) 
 \sim \frac{N^2}{4} - \frac{3N^\half}{2^4}E({\scriptstyle \frac 32}; \tau,\bar\tau)+\frac{45  N^{-\half}}{2^8}E({\scriptstyle \frac 52}; \tau,\bar\tau) \\
&\notag+  {N^{-\frac{3}{2}}}\Big[\frac{4725}{2^{15}} E({\scriptstyle \frac 72}; \tau,\bar\tau)-\frac{39}{2^{13}}E({\scriptstyle \frac 32}; \tau,\bar\tau) \Big] +{N^{-\frac{5}{2}}}\Big[ \frac{99225}{2^{18}} E({\scriptstyle \frac 92}; \tau,\bar\tau)  -\frac{1125}{2^{16}} E({\scriptstyle \frac 52}; \tau,\bar\tau)\Big]\\
&\notag +  {N^{-\frac{7}{2}}}\Big[\frac{245581875}{2^{27}}E({\scriptstyle \frac{11}{2}}; \tau,\bar\tau)  -\frac{2811375}{2^{25}}E({\scriptstyle \frac 72}; \tau,\bar\tau)+\frac{4599}{2^{22}} E({\scriptstyle \frac 32}; \tau,\bar\tau)\Big] \cr
&\notag +  {N^{-\frac{9}{2}}}\Big[\frac{29499294825}{2^{31}}E({\scriptstyle \frac{13}{2}}; \tau,\bar\tau)  -\frac{39590775 }{2^{26}}E({\scriptstyle \frac 92}; \tau,\bar\tau)+\frac{1548855 }{2^{27}} E({\scriptstyle \frac 52}; \tau,\bar\tau)\Big]  \cr
&\notag +  {N^{-\frac{11}{2}}}\Big[\frac{40266537436125}{2^{38}}E({\scriptstyle \frac{15}{2}}; \tau,\bar\tau)  -\frac{397105891875}{2^{36}}E({\scriptstyle \frac {11}{2} }; \tau,\bar\tau)+\frac{2029052025}{2^{34}} E({\scriptstyle \frac 72}; \tau,\bar\tau) \cr
&- \frac{3611751}{2^{32}} E({\scriptstyle \frac 32}; \tau,\bar\tau)\Big]  +O(N^{-\frac{13}{2}})  \,.
\end{align}
The corresponding expansions  for general classical gauge groups based on the large-$N$ expansion of  (\ref{eq:mainres}) were presented in \cite{Dorigoni:2022zcr}, and will not be reproduced here.

The first few terms in the large-$N$ expansion of  the second integrated correlator (\ref{eq:integrated2})  was presented in \cite{Chester:2020vyz}  and the result is 
\bea 
		& \nonumber \!\!\!\!  \!\!\!\!\!\!\!\!\!  \!\!\!\!\!\!\!\!\!  \!\!\!\!\!\!\!\!\! \!\!\!\!\! \partial_m^4 \log Z_{SU(N)}\big\vert_{m=0} \sim 6 N^2+ {6N^{\frac 12}}  E( {\scriptstyle {3 \over 2}};\tau,\bar\tau)+C_0-\frac{9}{2 N^{\frac 12} }E( {\scriptstyle {5 \over 2}};\tau,\bar\tau)-\frac{27}{2^3 N}\mathcal{E}({\scriptstyle {3 \over 2}},{\scriptstyle {3 \over 2}};3;\tau,\bar\tau)\\
		&  \!\!\!\!\!\!\!\!\!  \!\!\!\!\!\!\!\!\!  \!\!\!\!\!\!\!\!\!   \!\!\!\!\!\!\!\!\!  \!\!\!\!\!  \quad+\frac{9}{{N}^{\frac32}}\left[\frac{13}{2^8 }E( {\scriptstyle {3 \over 2}};\tau,\bar\tau)-\frac{375}{2^{10}  }E( {\scriptstyle {7 \over 2}};\tau,\bar\tau)\right]+\frac{405}{704N^2}\left[C_1+35 \mathcal{E}({\scriptstyle {5 \over 2}},{\scriptstyle {3 \over 2}};6;\tau,\bar\tau)- 24 \mathcal{E}({\scriptstyle {5 \over 2}},{\scriptstyle {3 \over 2}};4;\tau,\bar\tau)\right] \cr
		&  \!\!\!\!\!\!\!\!\!  \!\!\!\!\!\!\!\!\!  \!\!\!\!\!\!\!\!\!  \!\!\!\!\!\!\!\!\!  \!\!\!\!\!  \quad+\frac{675}{{N}^{\frac52}}\left[\frac{1}{2^{10} }E( {\scriptstyle {5 \over 2}};\tau,\bar\tau)-\frac{49}{2^{12} }E( {\scriptstyle {9 \over 2}};\tau,\bar\tau)\right]+\frac{1}{N^3}\Big[\alpha_3\, \mathcal{E}({\scriptstyle {3 \over 2}},{\scriptstyle {3 \over 2}};3;\tau,\bar\tau)\cr
		& \label{newpaper} \!\!\!\!\!\!\!\!\!  \!\!\!\!\!\!\!\!\!  \!\!\!\!\!\!\!\!\!  \!\!\!\!\!\!\!\!\!  \!\!\!\!\!  \quad +\sum_{r=5,7,9}[\alpha_r\, \mathcal{E}({\scriptstyle {3 \over 2}},{\scriptstyle {3 \over 2}};r;\tau,\bar\tau)+\beta_r\mathcal{E}({\scriptstyle {5 \over 2}},{\scriptstyle {5 \over 2}};r;\tau,\bar\tau)+\gamma_r\mathcal{E}({\scriptstyle {7 \over 2}},{\scriptstyle {3 \over 2}};r;\tau,\bar\tau)]\Big]+O(N^{-\frac72})\, , 
\eea
where the rational numbers  $\alpha_r, \beta_r$ and $\gamma_r$ can be found in \cite{Chester:2020vyz}.  The expansion in this case involves both half-integer and integer powers of  $1/N$ (whereas only half-integer powers appeared in  \eqref{oldexpand}).  Strikingly, the coefficients of the integer powers of $1/N$  involve sums of  non-holomorphic generalised Eisenstein series. The corresponding expansion for other classical gauge groups has not been determined.

The preceding two  expressions provide two constraints on the unknown coefficients in the  ansatz \eqref{eq:4ptsMst} for the unintegrated correlator at each order of the large-$N$ expansion.   Explicitly, these constraints lead to the values   
\bea
\tilde{f}_0 =2 \, , \qquad  f_1 = {15 E(\threeh; \tau,\bar{\tau}) \over 8 }\, , \qquad  f_{2,1}= -{1\over 3} f_{2,2} =  {315  E(\fiveh; \tau,\bar{\tau}) \over 128 }\, . 
\eea
Rather strikingly we see that terms in the low-energy expansion of type IIB superstring theory in an $AdS_5\times S^5$ background up to order $d^4R^4$  have been determined in this manner.

Furthermore, the ten-dimensional flat Minkowski-space limit can be obtained, according to \cite{Penedones:2010ue}, by taking the limit $s, t, u \rightarrow \infty$. In this limit (\ref{eq:4ptsMst}) reproduces precisely the same coefficients for the $R^4$ and $d^4R^4$ terms in  the low-energy expansion of type IIB superstring amplitudes as described in section~\ref{sec:4pts}.

However, two constraints from the integrated correlators (\ref{eq:integrated1}) and (\ref{eq:integrated2})  are not enough to fix the  three functions $f_{3,1}, f_{3,2}, f_{3,3}$ appearing at order of $N^{-1}$.  But if we include the flat-space result for the  $d^6R^4$ coefficient given in  (\ref{eq:LLeff})  as an input, together with the two constraints from the large-$N$ expansions (\ref{oldexpand}) and  (\ref{newpaper})  the unknown constants in the Mellin amplitude at this order are determined, and we find, 
\bea
f_{3,1}= -4 f_{3,2} =-{1\over 4} f_{3,3} =  {945 \mathcal{E}(\threeh, \threeh; 3; \tau) \over 32} \, .
\eea
Although this argument inputs the flat-space string theory $d^6R^4$ coefficient, it is  non-trivial that all three coefficients at order $N^{-1}$ in (\ref{eq:4ptsMst}) have been determined.  This adds further information concerning terms in the low-energy expansion of  the type IIB superstring action in $AdS_5\times S^5$.

Once the four-point correlator is determined, we can use equation \eqref{eq:recursion}, which is valid for all values of $N$ and $\tau$, to obtain information on MUV correlators. We may now construct higher-point MUV correlators recursively  using the  expression for the Mellin amplitude of the four-point correlator (\ref{eq:4ptsMst}) and solving the recursion relation order by order in the  large-$N$ expansion.   This leads to the large-$N$ expansion of $n$-point MUV correlator with finite coupling $\tau$   \cite{Green:2020eyj}. In the flat-space limit, the results at order $N^{1\over 2}$ and $N^{-{1\over 2}}$ again match with  the superstring low-energy expansion (\ref{eq:maximal-U(1)}) when  $p=0, 2$. As  with the four-point correlator, at order  $N^{-1}$ there are insufficient  constraints from integrated correlators to determine all the unknown coefficients in the low-energy expansion on $AdS_5\times S^5$.  Inputting the $p=3$ flat-space result from  (\ref{eq:p=3}) once again determines all the coefficients in the Mellin amplitude at this order.
 
  \setcounter{footnote}{0} 

 \section{Modular graph forms and superstring perturbation theory}
    \label{sec:otherasp}
 
In this section we will review another context in which modularity plays an important r\^ole.    The subject originated in the study of the low-energy expansion of superstring perturbation theory, but it has broader connections to areas of algebraic geometry and number theory.   Much of the literature focuses on genus-one modular graph forms \cite{Green:1999pv,Green:2008uj,DHoker:2015gmr,DHoker:2015wxz}, which are closely related to elliptic generalisations of multiple zeta-values, but there are obvious possible extensions to higher genus, which we have no space to cover in this article.

  This is a subject in which there have been many recent developments  both in the theoretical physics literature 
\cite{Green:1999pv, Green:2008uj, Green:2013bza, DHoker:2015gmr, DHoker:2015sve, Basu:2015ayg, DHoker:2015wxz, DHoker:2016mwo, Basu:2016xrt, Basu:2016kli, Basu:2016mmk, DHoker:2016quv, Kleinschmidt:2017ege, Basu:2017nhs, Broedel:2018izr, Ahlen:2018wng,Gerken:2018zcy, Gerken:2018jrq, DHoker:2019txf, Dorigoni:2019yoq, DHoker:2019xef, DHoker:2019mib, DHoker:2019blr, Basu:2019idd,  Gerken:2019cxz, Hohenegger:2019tii, Gerken:2020yii, Basu:2020kka, Basu:2020pey, Basu:2020iok , Hohenegger:2020slq}
and the mathematics literature \cite{Brown:mmv, Zerbini:2015rss, Brown:I, Brown:II, DHoker:2017zhq,Zerbini:2018sox, Zerbini:2018hgs, Zagier:2019eus, Berg:2019jhh}.   See also  the reviews \cite{Gerken:review} and \cite{Vanhove:2020qtt}, which cover much more of the literature than we can in this article, \cite{Gerken:2020aju} for a {\tt Mathematica} implementation and \cite{DHoker:2013fcx, DHoker:2014oxd, Pioline:2015qha, DHoker:2017pvk, DHoker:2018mys, Basu:2018bde, DHoker:2020tcq, DHoker:2020uid, Basu:2021xdt} for generalisations to higher genus.
 
   \subsection{Superstring perturbation theory} 
      
 \setcounter{footnote}{0}
In the  previous sections we have discussed $SL(2,\Z)$ S-duality of type IIB superstring theory and its connection with Montonen--Olive duality in $\cN =4$ SYM.  
However,  a different manifestation of modularity arises in string perturbation theory.  The perturbation expansion of a string theory amplitude  is a power series in $g_s^2$ in which a  term of order $g_s^{2g-2}$ is associated with a functional integral over a genus-$g$ world-sheet. For example, the perturbative expansion of a $n$-point massless scattering amplitude in ten-dimensional type IIB superstring theories  has the form
\begin{equation}
A^{(n)}(\epsilon_i, k_i; g_s) = \sum_{g=0}^\infty g_s^{2g-2} \mathcal{A}^{(n)}_g (\epsilon_i, k_i)\,,\label{eq:stringpert}
\end{equation}
where $(\epsilon_i,k_i)$  denotes  the polarisations and momenta of the scattered massless particles (with $i=1,2,\dots,n$).
  The Mandelstam variables  are defined by $s_{ij} = -\alpha'  (k_i + k_j )^2 /4$ as in section \ref{eq:superamp}. In the case of  the $4$-point function that is the main focus of  this section the standard convention is to define $s:=s_{12}= s_{34}\,,\,t :=s_{13}=s_{24}\,,\, u := s_{14}=s_{23}$,  which satisfy the condition $s+t+u =0$.   
Note that the perturbative part of the string amplitude  (\ref{eq:stringpert}) only depends on ${\rm Im}\, \tau= 1/g_s$, whereas the full amplitude has non-perturbative contributions, such as the contributions of D-instantons, that depend on $\tau= \tau_1+i \tau_2$.

The $g$-loop contribution, $\mathcal{A}^{(n)}_g (\epsilon_i, k_i)$, is, in principle,  given by a functional integral over all genus-$g$ (super) Riemann surfaces with $n$ punctures that represent the scattering particles. This includes integration over the fermionic supermoduli and summing over fermionic spin structures.   
However, there are technical obstacles in integrating over the fermionic moduli, which have so far prevented the explicit evaluation of  the four-graviton amplitude with $g\ge 3$.\footnote{The pure spinor formalism has no world-sheet spinors, but has other technical obstacles when $g\ge 3$.}  Nevertheless,  analysis of the degeneration limits of arbitrary genus super-Riemann surfaces has demonstrated the ultra-violet finiteness of these amplitudes.  For our  purposes it will be sufficient to restrict considerations to $g\le 2$, in which cases the resulting expressions  are expressed as integrals over the world-sheet  moduli, as well as over the positions of punctures on the world-sheet, that carry the information about the momenta and polarisations of the scattering particles.   

For concreteness, let us focus our attention to the well-studied case of the $10$-dimensional four-graviton scattering amplitude in type IIB superstring theory.
A general consequence of type IIB supersymmetry is that the four-graviton amplitude has a prefactor of $R^4$,  which denotes the particular scalar contraction of four  linearised Riemann curvature tensors, which we met in (\ref{eq:LLeff}).   This means that  the genus-$g$ term  in (\ref{eq:stringpert})  has the form 
\bea
 \mathcal{A}_g^{(4)}(\epsilon_i,k_i) =\kappa_{10}^2\, R^4 \, T_g(s,t,u)\,,
\label{genfour}
\eea
where $\kappa_{10}^2$ is  the ten-dimensional Newton constant, and $R^4$ (and its supersymmetric completion) may be expressed as $\delta^{16}(Q_4)$ using the spinor-helicity formalism given in section \ref{eq:superamp}. 
 The function $T_g(s,t,u)$ is a scalar function of the Mandelstam invariants that contains all of the non-trivial dynamical structure of the amplitude.
 
 The main emphasis in the following is the $g=1$ case, but  we will first very briefly review the structure of the four-graviton tree ($g=0$) amplitude, which  is determined by a functional integral over world-sheets of spherical topology and has the form
\begin{align}\label{eq:genus0}
\mathcal{A}_{g=0}^{(4)}(\epsilon,\,k_i) &= \frac{\kappa_{10}^2 R ^4}{stu}\frac{\Gamma(1-s)\Gamma(1-t)\Gamma(1-u)}{\Gamma(1+s)\Gamma(1+t)\Gamma(1+u)}\,,\\
&\notag =\frac{\kappa_{10}^2 R ^4}{stu}\exp\Big[\sum_{n=1}^\infty \frac{2\zeta(2n+1)}{2n+1} (s^{2n+1}+t^{2n+1}+u^{2n+1})\Big]\,.
\end{align}
The expression on the second line makes it obvious that the amplitude can be expanded as an infinite series of powers of  $s$, $t$ and $u$ with coefficients that are rational multiples of products of odd Riemann zeta-values.
It is also obvious that although products of zeta-values arise in this series there are no multiple zeta-values.

However, it has proved possible to analyse the expansion of tree-level $n$-point functions for all $n\in \N$ and to all orders in the low-energy expansion, see e.g. \cite{Schlotterer:2012ny}.  In the case of open superstrings this  is an expansion in monomials of Mandelstam invariants with rational coefficients multiplying multiple zeta-values, while in the case of closed  superstring amplitudes the coefficients are proportional to  single-valued multiple zeta-values.  These multiple zeta-values and single-valued multiple zeta-values appear when $n>4$, and the single-valued multiple zeta-values are defined as values of single-valued multiple polylogarithms  when its arguments are set equal to $1$ \cite{Brown:2004ugm,Brown:2013gia,Brown:sv}.

\subsection{The genus-one amplitude}

We will now focus on properties of the ($g=1$) four-point amplitude although our
considerations should generalise to an arbitrary amplitude for scattering of massless states at arbitrary order in the genus expansion.  A genus-one world-sheet, $\Sigma_\otau$, has the topology of a torus, which is diffeomorphic to $\mathbb{R}^2/\Lambda$, where the lattice $\Lambda = \otau \Z +\Z$ defines the shape of the torus.  This is parameterised by the complex structure, $\otau =\otau_1+i\otau_2$\footnote{In this section we are using a bold-faced symbol $\otau$  for the complex structure of the $g=1$ world-sheet in order not to confuse it with the complex scalar, $\tau$,  of ten-dimensional type IIB superstring.},  which takes values in the upper-half complex plane $\otau_2>0$, modulo discrete identifications that are associated with large diffeomorphisms associated with the modular group, $SL(2,\Z)$.

After performing the functional integral described above, the amplitude $\mathcal{A}_{g=1}^{(4)}(\epsilon_i\,, k_i)$ is expressed as an integral over the positions of the four punctures\footnote{Translation invariance means that there are only three integrals over the relative positions.} and  an integral over $\otau$  in a single fundamental domain of $SL(2,\Z)$.   This is often chosen, for convenience, to be   the domain $\mathcal{F}= \{ |\otau_1|\leq \frac{1}{2} \,,\,|\otau| \geq 1\}$,   After further integrating over the positions of the punctures the resulting $g=1$ amplitude takes the form\footnote{In writing (\ref{eq:genus1}) we have assumed that the integration over the positions of the punctures can be performed before the $\otau$ integral. However, this ignores the presence of branch cuts in $s,t,u$ that arise from the region $\otau_2\to \infty$.  For most of this section we are only interested in properties of the integrand $\mathcal{I}_4(s_{ij};\otau)$.  Where necessary we will cut off the $\otau$ integral at large $\otau_2$ in a consistent manner  \cite{Green:1999pv}.}
\begin{equation}
\mathcal{A}_{g=1}^{(4)}(\epsilon_i,k_i)  = 2\pi \kappa_{10}^2  R ^4\int_{\mathcal{F}} \frac{d^2\otau}{\otau_2^2} \mathcal{I}_4(s_{ij};\otau)\,
,\label{eq:genus1}
\end{equation}
where $\mathcal{I}_4(s_{ij};\otau)$ is a modular function that results from the integral
\begin{equation}
\mathcal{I}_4(s_{ij};\otau)= \int_{\Sigma_\otau} \Big( \prod_{i=2}^4 \frac{d^2 z_i}{\otau_2}\Big) \exp\Big( \sum_{1\leq i < j}^4 s_{ij} G(z_i -z_j | \otau) \Big)\,.
\label{eq:4grav}
\end{equation}
Here $G(z| \otau)$ denotes the scalar Green function on the torus $\Sigma_\otau$, which is defined to satisfy 
\bea
\label{3b0}
\Delta_z G(z|\otau) = - 4 \pi \delta ^{(2)} (z) + { 4 \pi \over \otau_2} \, ,  \qquad\quad \int _{\Sigma_\otau} d^2z \, G(z|\otau)=0\, ,
\eea
where  $\Delta_z= 4 \partial_{\bar z} \partial_z$, 
 which has the solution
\begin{equation}
G(z| \otau) = - \log \Big\vert \frac{\vartheta_1(z|\otau)}{\vartheta'_1(0|\otau)}\Big\vert^2 - \frac{\pi}{2\otau_2}(z-\bar{z})^2\,,
\end{equation}
and $\vartheta_1(z|\otau)$ is a Jacobi function. After changing from  $z$ to Cartesian coordinates on $\Sigma_\otau$, defined by   $z =u + v\otau$, the Green function can be written as the Fourier series
\begin{equation}
G(z|\otau) = \sum_{(m,n)\neq (0,0)}    \frac{\otau_2}{\pi |n \otau+m|^2} e^{2\pi i ( nu - m v)}\,,
\label{eq:FourierG}
\end{equation}  
where $(m,n)$ are integer-valued momentum components conjugate to $(v,u)$.

The genus-one four-point amplitude is a complicated non-analytic function  of $s_{ij}$.  This  non-analyticity  is indicated in integration over $\otau$ in  (\ref{eq:genus1}), which generates branch cuts associated with unitarity in  a well-understood fashion.  
However, its low-energy expansion can be treated by a systematic diagrammatic expansion \cite{Green:1999pv,Green:2008uj,DHoker:2015gmr,DHoker:2015wxz,DHoker:2016mwo}.  In  particular, the integrand $\mathcal{I}_4(s_{ij};\otau)$ can be expanded as a power series in $s$, $t$, $u$  by expanding the exponential term in \eqref{eq:4grav} (the  Koba-Nielsen factor), giving
\begin{equation}
\mathcal{I}_4(s_{ij};\otau) = \sum_{w=0}^\infty \frac{1}{w!} \int_{\Sigma_\otau} \Big( \prod_{i=2}^4 \frac{d^2 z_i}{\otau_2}\Big) \Big(\sum_{1\leq i < j \leq 4} s_{ij} G(z_i-z_j \vert \otau)\Big)^w\,.\label{eq:GreenExp}
\end{equation}
This expression is a contribution of order $(\alpha')^w$ to the low-energy expansion.  This means it is a contribution  to the type IIB effective action of order $d^{2w} R^4$ where the pattern of contractions of the derivatives is specified by the powers of $s$, $t$ and $u$.  More precisely, given that $\mathcal{I}_4(s,t,u;\otau)$ in \eqref{eq:GreenExp} is a symmetric function of $s$, $t$, $u$, subject to $s+t+u=0$.  This means that, in the $\alpha'$ expansion, $\mathcal{I}_4(s,t,u;\otau)$ can be expressed 
\begin{equation}
\mathcal{I}_4(s,t,u;\otau) = \sum_{p,q\geq0} F_{p,q} (\otau) (s^2+t^2+u^2)^p(s^3+t^3+u^3)^q \, ,
\end{equation} 
where $2p+3q=w$ and the coefficients $F_{p,q}(\otau)$ are sums  of modular objects called modular graph functions for reasons that will shortly be clear.

Each coefficient $F_{p,q}(\otau)$ is the sum of terms in which $w$ Green functions join pairs of points labelled by the $z_i$.   This  can be represented by a sum of Feynman diagrams on the two-torus. Each diagram consists of the product of Green functions joining pairs of points at positions $z_i$ and $z_j$, which are integrated over $\Sigma_\otau$. 
It is convenient to represent the diagram in  momentum space, where the momentum-space propagator is given by $\otau_2/(\pi\, |p|^2)$, with  $p=n\otau +m\in \Lambda\setminus\{0\}$, as follows from the expression for the Green function in (\ref{eq:FourierG}).

A general diagram has $\ell_{ij}$ propagators joining any pair of vertices labelled $i$ and $j$, where $i,j\in\{1,2,3,4\}$. There are therefore  six $(i,j)$ pairs so a general diagram has $\ell_1, \dots,\, \ell_6$ propagators.   This  notation is summarized by the following diagram:
 \vskip3pt
 \begin{center}
 \tikzpicture[scale=1.9]
\begin{scope}[xshift=6cm]
\draw (0,0) node{$\bullet$} ;
\draw (1,0) node{$\bullet$} ;
\draw (0,1) node{$\bullet$} ;
\draw (1,1) node{$\bullet$} ;
\draw (0,0) -- (1,0) ;
\draw (0,0) -- (0,1);
\draw (1,1) -- (1,0) ;
\draw (1,1) -- (0,1);
\draw (0,0) -- (1,1) ;
\draw (0,1) -- (1,0) ;
\draw (0,0.5) [fill=white] circle(0.15cm) ;
\draw (0,0.5)  node{$\ell_1$};
\draw (0.5,0) [fill=white] circle(0.15cm) ;
\draw (0.5,0)  node{$\ell_2$};
\draw (1,0.5) [fill=white] circle(0.15cm) ;
\draw (1,0.5)  node{$\ell_3$};
\draw (0.5,1) [fill=white] circle(0.15cm) ;
\draw (0.5,1)  node{$\ell_4$};
\draw (0.3,0.3) [fill=white] circle(0.15cm) ;
\draw (0.3,0.3)  node{$\ell_5$};
\draw (0.3,0.7) [fill=white] circle(0.15cm) ;
\draw (0.3,0.7)  node{$\ell_6$};
\draw(2,0.5) node{$= \ D_{\ell_1,\ell_2,\ell_3,\ell_4;\ell_5,\ell_6}$};
\end{scope}
\endtikzpicture
 \end{center}
 \vskip3pt
where the label $\ell$ on the link  
\begin{tikzpicture} [scale=1.3]
\scope[xshift=0cm,yshift=0cm]
\draw (0,-0.0) node{$\bullet$}   ;
\draw (1,0.0) node{$\bullet$}  ;
\draw (0,0) -- (1,0) ;
\draw (0.5,0) [fill=white] circle(0.15cm) ;
\draw (0.5,0.0) node{$\ell$};
\endscope
\end{tikzpicture}
indicates the product of $\ell$ propagators joining the corresponding pair of vertices. This diagram is a tetrahedron that is symmetric in all edges and vertices and its {\it weight}, $w$,  is given by the number of propagators, so that  $w= \sum_{r =1}^6 \ell_r$,  where $0\le \ell_r $.\footnote{This weight should not be confused with the modular weight, which vanishes for the modular-graph functions.}     When                                                                                                                                                                                                                                                                                                                                                                                                                                                                                                                                                                                                                                                                                                                                                                                                                                                                                                                                                                                                                                                                                                                                                                                                                                                                                                                                                                                   $\ell_r =0$ for particular values of $r$,  the diagram degenerates to a simpler diagram. In such cases, we shall omit the index that vanishes.\footnote{For example $D_{\ell_1,\ell_2,\ell_3,\ell_4;0,0}= D_{\ell_1,\ell_2,\ell_3,\ell_4}$, while $ D_{\ell,0,0,0,;0,0} = D_{\ell}$, etc.}

The integral over the positions of the vertices, $z_i$ can be easily performed thanks to the Fourier series representation \eqref{eq:FourierG},  leading to the general momentum-space expression for a \textit{modular graph function:}
\begin{equation}
\label{gengraph}
I_\Gamma(\otau,\bar\otau)=  \sum_{p_1,\dots,p_w\in\Lambda}' ~ \prod_{\alpha =1}^w
  {\otau_2\over  \pi|p_\alpha |^2}\, \prod_{i =1}^N
  \delta \left ( \sum_{\alpha =1}^w C_{i \alpha}  p_\alpha \right )\,,
\end{equation}
where  $\Gamma$ denotes the connectivity of the graph and we have generalised  the rules  to include  $N$ vertices (where $N$ is an arbitrary integer) even though only four vertices are involved in the expansion of the four-particle amplitude. In this expression $p_\alpha =n_\alpha \otau + m_\alpha\in\Lambda$ labels the momentum in the link labelled $\alpha$ and the prime above the summation symbol indicates that the sums over $p_\alpha$ exclude the value 0. The Kronecker $\delta$ symbol enforces momentum conservation at each vertex -- it takes the value 1 when its argument vanishes and zero otherwise; the coefficients  
$C_{i \alpha}$ are given as follows
\bea
C_{i \alpha} = \left \{ 
\begin{matrix} 
\pm 1 & \hbox{if edge $\alpha$ ends on vertex $i$} \\ & \\
0  & \hbox{otherwise} 
\end{matrix} 
\right .
\eea 
the sign being determined by the orientation of the momenta.

We stress that the expression (\ref{gengraph})  is a multiple sum that generalises the non-holomorphic Eisenstein series and  is manifestly invariant under $SL(2,\Z)$ transformations acting on $\otau$.

Although the analysis of the properties of general modular graph functions is presently rather rudimentary we turn now to consider a special class of such functions about which a great deal is known.  
These are modular graph functions that are defined by graphs that have one or two loops and any number of vertices.  

\subsection*{One-loop modular graph functions}

Since the zero mode of the Green function vanishes \eqref{3b0}, it is obvious that a graph has to consist of closed loops of propagators.  The simplest class of such functions is therefore  represented by one-loop graphs.
A one-loop modular graph function with $a$ vertices is represented by

\vskip3pt 
\begin{center}
\tikzpicture[scale=1.6]
\scope[xshift=-0.0cm,yshift=-0.0cm]
\draw (0,0.0) node{$\bullet$}   ;
\draw (1,0) node{$\bullet$} ;
\draw (1,0) arc[x radius=0.5, y radius=0.5, start angle=0, end angle=-230];
\draw [dashed] (1,0) arc[x radius=0.5, y radius=0.5, start angle=0, end angle=130];
\draw (0.5,0.50)node{$\bullet$} ;
\draw (0.5,-0.50) node{$\bullet$} ;
\draw (0.14644,-0.35356)node{$\bullet$} ;
\draw (0.14644, 0.35356)node{$\bullet$} ;
\draw (0.85355, - 0.35356)node{$\bullet$} ;
\draw (0.85355, 0.35356)node{$\bullet$} ;
\draw(3,0.) node{$= \ E(a; \otau, \bar \otau) = \sum^\prime_{p \in \Lambda}  \left(\frac{ \otau_2}{\pi |p|^2} \right)^a $} ;
\endscope
\endtikzpicture
\end{center}
\vskip3pt
and is simply a non-holomorphic Eisenstein series.

  \subsection{Two-loop modular graph functions}

The two-loop  modular graph functions are much less familiar so we will discuss their properties in some detail.  They are  represented by graphs with three chains and are denoted $C_{a,b,c}(\otau,\bar\otau)$.
 They are represented pictorially by 
\vskip3pt 
\begin{center}
\tikzpicture[scale=1.7]
\scope[xshift=-5cm,yshift=-0.4cm]
\draw (0,0.0) node{$\bullet$}   ;
\draw (1,0) node{$\bullet$} ;
\draw (0,0) -- (1,0) ;
\draw (0.5,0.0) ellipse (0.5 and 0.3);
\draw (0.375,0.175) [fill=white] rectangle (0.625,0.425) ;
\draw (0.5,0.30) node{$a$};
\draw (0.375,-0.125) [fill=white] rectangle (0.625,0.125) ;
\draw (0.5,0) node{$b$};
\draw (0.375,-0.425) [fill=white] rectangle (0.625,-0.175) ;
\draw (0.5,-0.30) node{$c$};
\draw (1.5,0) node{$=$};
\draw (1.9,0) node{$\bullet$} ;
\draw (2.4,0) node{$\bullet$} ;
\draw (2.9,0) node{$\bullet$} ;
\draw (3.4,0) node{$\bullet$} ;
\draw (3.9,0) node{$\bullet$} ;
\draw (1.9,0) -- (2.4,0) ;
\draw (2.4,0) -- (2.9,0) ;
\draw[dashed, thick] (2.9,0) -- (3.4,0) ;
\draw (3.4,0) -- (3.9,0) ;
\draw (2.4,0.25) node{$\bullet$} ;
\draw (2.9,0.30) node{$\bullet$} ;
\draw (3.4,0.25) node{$\bullet$} ;
\draw (1.9,0) -- (2.4,0.25) ;
\draw (2.4,0.25) -- (2.9,0.3) ;
\draw[dashed, thick] (2.9,0.3) -- (3.4,0.25) ;
\draw (3.4,0.25) -- (3.9,0) ;
\draw (2.4,-0.25) node{$\bullet$} ;
\draw (2.9,-0.30) node{$\bullet$} ;
\draw (3.4,-0.25) node{$\bullet$} ;
\draw (1.9,0) -- (2.4,-0.25) ;
\draw (2.4,-0.25) -- (2.9,-0.3) ;
\draw[dashed, thick] (2.9,-0.3) -- (3.4,-0.25) ;
\draw (3.4,-0.25) -- (3.9,0) ;
\draw(5.0,0) node{$= \  C_{a,b,c}(\otau, \bar \otau)$}; 
\endscope
\endtikzpicture 
\end{center}
\vskip3pt
The integers $a$, $b$, $c$ are the number of propagators that are joined end to end along each chain.

It follows from the preceding graphical rules that the expression for the function represented by the above graph is 
 \begin{equation}
C_{a,b,c}(\otau,\bar{\otau})  = \Big(\frac{\otau_2}{\pi}\Big)^{a+b+c} \sum^\prime_{p_1,p_2,p_3 \in \Lambda} \frac{\delta(p_1+p_2+p_3)}{|p_1|^{2a} |p_2|^{2b} |p_3|^{2c}}\,,\label{eq:Cabc}
\end{equation}
with $a,b,c\in\mathbb{N}$.
In \cite{DHoker:2015gmr} it was shown that the $C_{a,b,c}$ satisfy a closed system of inhomogeneous Laplace equations. 
The simplest example, $C_{1,1,1}(\otau,\bar{\otau})$ has weight $w=3$ and contributes to the $d^6R^4$ term in low-energy expansion of the genus-one amplitude \cite{Green:1999pv}.  It satisfies the equation 
\bea
\label{eq:c111}
\Delta_\otau C_{1,1,1}(\otau,\bar{\otau}) = 6 E(3; \otau,\bar{\otau})\,,
\eea
which has solution $C_{1,1,1}(\otau,\bar{\otau}) = E(3; \otau,\bar{\otau}) + \zeta(3)$ where the constant  $\zeta(3)$ is determined by a boundary condition that is obtained by computing the asymptotic behaviour of the lattice sum \eqref{eq:Cabc} at the cusp $\otau_2\to \infty$ (as in \cite{DHoker:2015gmr}). It can also be calculated directly from the lattice sum \cite{Zagier}.

Although our primary interest in this section is in modular properties of the integrand
 $\mathcal{I}_4(s_{ij};\otau)\,$ at this point we will comment on the evaluation of its $\otau$ integral in  (\ref{eq:genus1}).
From \eqref{eq:GreenExp}, one can easily see that the complete contribution at this order is
\begin{equation}
\frac{1}{3!} \int_{\Sigma_\otau} \Big( \prod_{i=2}^4 \frac{d^2 z_i}{\otau_2}\Big) \Big(\sum_{1\leq i < j \leq 4} s_{ij} G(z_i-z_j \vert \otau)\Big)^3 = \frac{s^3+t^3+u^3 \,}{3!} \Big[ 8 E(3;\otau,\bar{\otau}) + 2 C_{1,1,1}(\otau,\bar{\otau})\Big]\,.
\end{equation}
Making use of $C_{1,1,1}(\otau,\bar{\otau}) = E(3; \otau,\bar{\otau}) + \zeta(3)$, and after integrating over the fundamental domain $\mathcal{F}$  \cite{Green:1999pv}\footnote{More precisely,  since the integral diverges a cut-off is introduced at  $\otau_2 = L\gg 1$  and the integral is restricted to the cut-off fundamental domain, $\cF_L$. The dependence on $L$ cancels after careful analysis of the non-analytic threshold contributions, so we have effectively  $\int_{\cF} d^2\otau/\otau_2^2 \,  E(s;\otau,\bar\otau)\sim 0$ (which is in accord with the mathematical observations in \cite{Zagierb}).}, the value of the genus-one  contribution at order $d^6 R ^4$  is  found to be
\begin{equation}
{4\over 3} \zeta(2)\zeta(3)\, (s^3+t^3+u^3)\,.
\end{equation}
This is the precise value  contained in the zero mode of the  non-perturbative $d^6R^4$ coefficient function  $g_s\,\mathcal{E}(\threeh,\threeh;3;\tau,\bar{\tau})$ that was derived in  \cite{Green:2005ba} (and is displayed in \eqref{eq:ME-per}).
 
Another simple example is
\bea
\Delta_\otau C_{2,2,1}(\otau,\bar{\otau}) = 8 E(5; \otau,\bar{\otau})\,,
\label{eq::c221}
\eea
with solution $C_{2,2,1} (\otau,\bar{\otau})= 2 E(5; \otau,\bar{\otau}) /5+ \zeta(5)/30$.

More complicated examples have source terms that are quadratic in non-holomorphic Eisenstein series.  For example, 
\begin{align}
(\Delta_\otau -2) C_{2,1,1}(\otau,\bar{\otau}) &\label{eq:LapC211}= 9 E(4; \otau,\bar{\otau}) - E(2; \otau,\bar{\otau})^2\,,\\
(\Delta_\otau - 6) C_{3,1,1}(\otau,\bar{\otau}) &\label{eq:LapC311} = \frac{86}{5} E(5; \otau,\bar{\otau})-4 E(2; \otau,\bar{\otau})E(3; \otau,\bar{\otau})+\frac{\zeta(5)}{10}\,.
\end{align}
Both of these examples arise in the low-energy expansion of the four-point amplitude, at order $d^8R^4$ and $d^{10}R^4$, respectively.

More generally, in \cite{Gerken:2019cxz,Gerken:2020yii} a generating series was  introduced, to produce all integrals over world-sheet tori which appear in closed-string one-loop amplitudes. Using these results, in \cite{Dorigoni:2021jfr,Dorigoni:2021ngn} it was proved that all depth-two\footnote{Here the {\it depth} of a modular graph function is defined to be the maximum depth of the multiple zeta values that arise as coefficients of powers of $\otau_2$ in its Laurent polynomial.} $C_{a,b,c}$ can be obtained from the generalised Eisenstein series 
\begin{equation}
(\Delta_\otau - r(r+1)) \mathcal{E}(m,k;r;\otau,\bar{\otau}) = -E(m;\otau,\bar{\otau}) E(k;\otau,\bar{\otau})\,,\label{eq:GenEisC}
\end{equation}
with $m,k \in \mathbb{N}$ and $m,k\geq 2$, and spectrum $r\in\{|k-m|+1,|k-m|+3, \ldots, k+m-5,k+m-3\}$, where the generalised Eisenstein $\mathcal{E}(m,k;r;\otau,\bar{\otau}) $ corresponds to $-{\rm F}^{+\,(r-1)}_{m,k}$ in those references.
In particular, it was shown that any $C_{a,b,c}$ with weight $w=a+b+c$, is given by rational linear combinations of finitely many $ \mathcal{E}(m,k;r ;\otau,\bar{\otau})$ with $w=k+m$, modulo the addition of a rational multiple of a non-holomorphic Eisenstein $E(w;\otau,\bar{\otau})$ and, in the case of odd weight $w$, a  rational multiple of $\zeta(w)$.

For example we have
\begin{align}
C_{2,1,1}(\otau,\bar{\otau}) & = \mathcal{E}(2,2;1;\otau,\bar{\otau}) +\frac{9}{10} E(4;\otau,\bar{\otau})\,,\\
C_{3,1,1}(\otau,\bar{\otau}) &= 4\, \mathcal{E}(2,3;2;\otau,\bar{\otau}) +\frac{43}{35} E(5;\otau,\bar{\otau}) -\frac{\zeta(5)}{60}\,,
\end{align}
which can be shown to be consistent with the Laplace equations \eqref{eq:LapC211}-\eqref{eq:LapC311}.

We should stress that, generically, the space of generalised Eisenstein series, $\mathcal{E}(m,k;r;\otau,\bar{\otau})$, defined above is larger that the space of $C_{a,b,c}$ (modulo single Eisenstein series and constant terms).  We now compare  the dimensions of the vector space of generalised Eisenstein series \eqref{eq:GenEisC} with the dimension of the vector space of $C_{a,b,c}$ (modulo single Eisenstein series and constants).  Denoting the dimensions of these spaces by $\mbox{dim}\mathcal{V}_{\mathcal{E}}(w,r)$ and $\mbox{dim}\mathcal{V}_{C}(w,r)$, respectively, where the weights are fixed to be $w=k+m = a+b+c$ and the eigenvalues are  $r(r+1)$, we have \cite{Dorigoni:2021jfr}
\begin{equation}
\dim \mathcal{V}_{\mathcal{E}}(w,r)- \dim \mathcal{V}_C(w,r) = \dim \mathcal{S}_{2(r+1)} = \left\lbrace \begin{array}{lc}
\left\lfloor \frac{2(r+1)}{12}\right\rfloor -1\,\,\, :& 2(r+1)\equiv 2 \!\!\!\mod 12\,,\\[2mm]
 \left\lfloor \frac{2(r+1)}{12}\right\rfloor \phantom{-1}\,\,\,\,\, :& \mbox{otherwise}
\end{array}\right.
\label{dimsmatch}
\end{equation}
where $ \mathcal{S}_{2(r+1)} $ denotes the vector space of holomorphic cusp forms with modular weight $2(r+1)$.

This appearance of holomorphic cusp forms is not a mere coincidence. In \cite{Dorigoni:2021ngn} it was  shown
that special completed L-values of holomorphic cusp forms appear in the non-zero modes  of the  Fourier series decomposition of $\mathcal{E}(m,k;r;\otau,\bar{\otau})$ with respect to $\otau_1$  precisely when  $\mbox{dim}\mathcal{V}_{\mathcal{E}}(w,r)>\mbox{dim}\mathcal{V}_{C}(w,r)$.  They arise as a consequence of modularity.

This mixing with holomorphic cusp forms becomes more manifest once we choose to represent the generalised Eisenstein series as particular combinations of iterated integrals of holomorphic Eisenstein series \cite{Broedel:2014vla,Broedel:2015hia,Broedel:2018izr,Dorigoni:2021jfr,Dorigoni:2021ngn}.  The Eichler-Shimura theorem \cite{Eichler:1957,Shimura:1959} and the work of Brown \cite{Brown:I,Brown:II,Brown:III,Brown:mmv} on iterated integrals of general holomorphic modular forms makes it very plausible that holomorphic cusp forms make an appearance.

However, the generating series \cite{Gerken:2019cxz,Gerken:2020yii} for modular forms arising in closed-string one-loop amplitudes contains conjectural matrix representations of Tsunogai's derivation algebra \cite{Tsunogai}. Relations in this algebra are known to be related to holomorphic cusp forms \cite{Pollack}, and, precisely due to these special selection rules governed by Tsunogai's derivation algebra, only the combinations of generalised Eisensteins series for which the cusp forms drop out are the ones appearing in the generating series for all the building blocks of one-loop type II superstring amplitudes.

Furthermore,  knowing how to decompose modular graph functions into a basis of modular objects satisfying inhomogeneous Laplace equations, such as \eqref{eq:GenEisC}, is useful for specific calculations, such as  the evaluation  of the integral \eqref{eq:genus1} over the modular parameter $\otau$ on the fundamental domain $\mathcal{F}$.

General results for the Laurent polynomials of $C_{a,b,c}$, i.e. the perturbative expansion at the cusp in the zero-Fourier mode sector, where obtained in  \cite{DHoker:2017zhq} starting from the lattice sum representation \eqref{eq:Cabc}. Similarly, in \cite{Ahlen:2018wng,DHoker:2019txf,Dorigoni:2019yoq} a Poincar\'e series approach was used to obtain consistent expressions.  The complete asymptotic behaviour of  $\mathcal{E}(m,k;r;\otau,\bar{\otau})$ with $m,k \in \mathbb{N}$ and $m,k\geq 2$, and $r\in\{|k-m|+1,|k-m|+3, \ldots, k+m-5,k+m-3\}$ was derived in \cite{Dorigoni:2021jfr}, making use of the Poincar\'e  series.
In general both $C_{a,b,c}(\otau,\bar{\otau})$ and  $\mathcal{E}(m,k;r;\otau,\bar{\otau})$ have a Laurent polynomial consistent with uniform trascendentality, meaning that if we assign trascendentality $1$ to $y =\pi\otau_2$ and trascendentality $k$ to $\zeta(k)$, then each monomial in the Laurent polynomials of both $C_{a,b,c}(\otau,\bar{\otau})$ and  $\mathcal{E}(m,k;r;\otau,\bar{\otau})$ has trascendentality $w=a+b+c= k+m$.
For example,
\begin{align}
C_{2,1,1}(\otau,\bar{\otau}) & = \frac{2y^4}{14175}+\frac{\zeta(3) y}{45} +\frac{5 \zeta(5)}{12y}- \frac{\zeta(3)^2}{4y^2} + \frac{9\zeta(7)}{16y^3} + O(q,\bar{q})\,,\\
C_{3,1,1}(\otau,\bar{\otau}) & = \frac{2y^5}{155925} + \frac{2\zeta(3) y^2}{945} - \frac{\zeta(5)}{180} +\frac{7\zeta(7)}{16 y^2} - \frac{\zeta(3)\zeta(5)}{2y^3} +\frac{43 \zeta(9)}{64y^4}+O(q,\bar{q})\, ,
\end{align}
where $q=e^{2\pi i \,\otau},  \, \bar q=e^{ - 2\pi i \, \bar \otau}$. Therefore $O(q,\bar{q})$ represents exponentially decaying  terms.

\subsection{Some further comments concerning genus-one modular graph functions}

We here  note some further general points that we have no space to elaborate on. 

\begin{itemize}

\item
  Modular graph functions satisfy a host of very impressive identities \cite{DHoker:2015sve, DHoker:2015gmr}. These are analogous to identities that relate multiple zeta-values of a given weight.  Among the many such relationships that have been discovered  are several that have proved important for  evaluating the coefficients in the low-energy expansion of the genus-one four-point amplitude.    These coefficients  involve integration of a combination of modular graph functions over $\otau$.  
  
Here we will simply mention two of these identities.  The following weight-$4$ identity is important for evaluating the  coefficient at order $d^8R^4$:
 \bea
 \label{eq:D4iden}
 D_4(\otau,\bar\otau) =24 C_{2,1,1}(\otau,\bar\otau) - 18 E(4;\otau,\bar\otau) + 3 E(2;\otau,\bar\otau)^2\, ,
 \eea
 where, following the footnote preceding (\ref{gengraph}),  the modular graph function $D_4\equiv D_{4,0,0,0;0,0}$.
Expressing $D_4$ in terms of $C_{2,1,1}$ and Eisenstein series  in this way  is the key to integrating over $\otau$ that is necessary for evaluating  the coefficients in the low-energy expansion at order $d^8R^4$.

The evaluation of the coefficient of order $d^{10}R^4$ makes use of several highly non-trivial weight-5 identities.  One of these is 
  \bea
\!\!\!\!\!\!\!\!\!\!\!\!\!\!\!\!\!\!  \!\!\!\!\!\! \!\!\!\!\!\! \!\!\!
  D_5(\otau,\bar\otau) = 60 \, C_{3,1,1}(\otau,\bar\otau)+10\, E(2; \otau,\bar\otau)\, C_{1,1,1}(\otau,\bar\otau)-48\, E(5; \otau,\bar\otau)   +16 \zeta(5)\,.
  \label{eq:D5den}
  \eea
 This relation between $D_5$ and $C_{3,1,1}$, together with other identities that relate $D_{3,1,1}$ and $D_{2,2,1}$ to $C_{3,1,1}$ and Eisenstein series again provide the basis for evaluating the integral over $\otau$.

\item
The Laurent polynomial of  a modular graph function of weight $w$   (the zero Fourier mode)   is a series of terms with integer  powers of $y=\pi \otau_2$ ranging from $y^w$ to $y^{1-w}$. The coefficients of the terms in this series were argued in  \cite{DHoker:2015wxz, Zerbini:2015rss}  to be rational  multiples of single-valued  multiple zeta-values. 
 The first example of an irreducible multiple zeta value arising as a coefficient in a Laurent series  was found in \cite{Zerbini:2015rss} where  the coefficient of the $y^{-4}$ term in the  Laurent polynomial of $D_{5,1,1}$ was found to be the weight-11 single-valued multiple zeta-value,
\bea
\label{singval}
\zeta^{sv}(3,5,3) = 2 \zeta(3,5,3)-2\zeta(3) \zeta(3,5)-10 \zeta(3)^2 \zeta(5)\,,
\eea 
and $\zeta(i,j)$ and $\zeta(i,j,k)$ are non single-valued multiple zeta-values.

\item
Modular graph functions are related to  elliptical  generalisations of  single-valued multiple polylogarithms in much the same way as as single-valued   multiple zeta-values are related to single-valued   multiple polylogarithms \cite{DHoker:2015wxz}.

\item
The expression (\ref{gengraph}) does not describe the most general modular graph functions that contribute to  the low-energy expansion of the $n$-point amplitude when $n>4$. The more general contributions that first enter at $n=5$  \cite{Green:2013bza} are modular functions in which there are `holomorphic' propagators of the form $\otau_2/p$ and anti-holomorphic propagators of the form $\otau_2/ \bar p$. In order for the total modular weight to vanish there must be equal numbers of holomorphic and anti-holomorphic propagators in any graph.  

 \item
 More generally still,  in considering  relationships between modular graph functions it is important to include modular graph forms, which transform with non-zero total modular weight.  These have unequal numbers of holomorphic and anti-holomorphic propagators  and are  related to each other by multiple applications of the Cauchy--Riemann operator \cite{DHoker:2016mwo,DHoker:2016quv}.

\end{itemize} 

\subsection{Genus-two modular graph functions}

Much less is known about higher-genus modular graph functions.    
 The genus-two four-point amplitude  in type II superstring theory was evaluated explicitly in the Ramond--Neveu--Schwarz formalism in \cite{DHoker:2001kkt, DHoker:2005vch}   and later in the pure spinor formulation  in  
  \cite{Berkovits:2005df,Berkovits:2005ng}.
  It is given by an integral over the moduli space $\cM_2 \approx  \cH_2/Sp(4,\ZZ)$ of genus-two Riemann surfaces $\Sigma $, where $\cH_2$ is the Siegel upper half space, which is parameterised  by the $2\times 2$ period matrix, $\Omega$. As at genus-one the integrand is an integral  over four points on $\Sigma$,  corresponding to the four gravitons. The low-energy expansion of the integral over the points, without integrating over $\cM_2$, now gives rise to $Sp(4,\Z)$-invariant functions, which are genus-two modular graph functions \cite{DHoker:2017pvk}.   
  
  It is interesting to consider the behaviour  of these functions in the limit that a handle on the genus-two world-sheet degenerates.  The  non-separating degeneration can be parameterised by a suitably chosen real variable $t$.  In the limit   $t \to \infty$   the surface reduces to a genus-one surface with two marked points separated by a distance $v$ in suitable coordinates (in addition to the four that correspond to the external particles). 
 In the degeneration limit, a genus-two modular graph function has the form of a Laurent polynomial in $t$ with exponentially small corrections
\bea
\label{LaurentBpq}
\cZ_i  (\Omega) = \sum_{n=-w}^{w} (\pi t)^n \, \mz_i^{(n)}(v|\otau)  + \cO(e^{-2\pi t})\,,
\eea 
where $\otau$ is the complex structure of the residual torus.  The coefficients $\mz_i^{(n)}(v|\otau)$ are  non-holomorphic Jacobi forms, which are elliptic functions closely related to genus-one modular graph functions.  This is reminiscent of the pattern of coefficients of powers of $(\pi \,\otau_2)$  in the Laurent expansion of genus-one modular graph functions, which are multiple zeta-values.

The low-energy expansion of the two-loop amplitude starts with the  effective interaction $d^4R^4$ with a constant coefficient that is proportional to the volume of  genus-two moduli space.  The value of this constant  matches the predictions of S-duality in Type~IIB string theory \cite{DHoker:2005jhf}  which comes from the genus-two term in the zero mode of $E(\fiveh;\tau,\bar\tau)$ in (\ref{eq:LLeff}). The next term in the low-energy expansion is $d^6 R^4$, which is obtained by bringing down a single Green function inside the genus-two  integrand.  
Its coefficient is a non-trivial $Sp(4,\ZZ)$-invariant function \cite{DHoker:2013fcx}, known as the Kawazumi--Zhang invariant, which satisfies a  Laplace eigenvalue equation on $\cH_2$  \cite{DHoker:2014oxd},  and which has an elegant representation as a generalised theta-lift \cite{Pioline:2015qha}.  Its integral over $\cM_2$ was computed using the Laplace equation and also matches a prediction of  S-duality, which is given by the coefficient of the genus-two term  in the zero mode of $\cE(\threeh,\threeh;3;\tau,\bar\tau) $ in (\ref{eq:LLeff}).

The next order  in the low-energy expansion involves integrating  the product of two Green functions and contributes the genus-two coefficient of the $d^8R^4$ interaction.  Detailed  properties of this modular graph function may be found  in \cite{DHoker:2018mys}. However, its integral over genus-two moduli space has not been carried out yet.  

There are no explicit expressions for type II superstring loop amplitudes of genus higher than two although an impressive calculation \cite{Gomez:2013sla} determined the leading low-energy behaviour of the  genus-three four-point amplitude, which is of order $d^6R^4$. Its value again agrees with the S-duality prediction, which is the coefficient of the genus-three component of the zero Fourier mode of $\cE(\threeh,\threeh;3;\tau,\bar\tau) $ in (\ref{eq:LLeff}). Further analysis of the genus-three amplitude  is given in  \cite{Geyer:2021oox}.

\section{Coda}
\label{sec:conclude}
 
This article has surveyed recent developments in three interrelated areas of string theory and quantum field theory that have several  themes in common.  These topics all involve the strong constraints of maximal supersymmetry -- ten-dimensional type IIB in the context of the superstring discussions in sections \ref{sec:susymod}   and  \ref{sec:otherasp}, and four-dimensional $\cN=4$ supersymmetry and superconformal symmetry in the context of integrated correlators in section~\ref{sec:correlator}. Another common theme is that of the strong constraints imposed by duality, which is target-space $SL(2,\Z)$  invariance of type IIB superstring, Montonen--Olive duality in the case of  $\cN=4$ SYM, and world-sheet duality in the case of genus-one or genus-two string perturbation theory.  

These constraints  are so strong that they lead to remarkably detailed expressions in each of these areas.  These results  not only shed light on areas of direct interest in theoretical physics, but they have led to interesting avenues   of significant mathematical interest.   However,  we have focussed on particular special examples and it would be interesting to  extend the ideas and methods covered  in this article to more general physical observables as well as more general systems.

\section*{Acknowledgments}
This work was supported by the European Union's Horizon 2020 research
and innovation programme under the Marie Sk\l{}odowska-Curie grant
agreement No.~764850 {\it ``\href{https://sagex.org}{SAGEX}''}. MBG has been partially supported by STFC consolidated grant ST/L000385/1. CW is supported by a Royal Society University Research Fellowship No. UF160350.\\

\appendix

\section{Some  properties of non-holomorphic modular forms}
\label{sec:math}

In this appendix we will briefly review the mathematics of non-holomorphic modular forms. Recall that $SL(2,\ZZ)$ acts on the scalar field $\tau = \tau_1 + i\tau_2$ (or equivalently on the worldsheet torus complex structure $\otau$ in section \ref{sec:otherasp}) as
\begin{equation}
\label{tautrans}
\tau\to \gamma\cdot\tau =  \frac{a\tau+ b}{c\tau+d}\,,
\end{equation}
 with $\gamma = \left(\begin{smallmatrix} a & b \\ c & d \end{smallmatrix}\right)\in SL(2,\ZZ)$ so that $a,b,c,d\in \ZZ$ and $\mbox{det}\,\gamma = ad-bc=1$.  

 An element $f^{(w,\hat w)}(\tau,\bar{\tau})$ of the vector space $M_{w,\hat{w}}$ of non-holomorphic modular forms with holomorphic and anti-holomorphic modular weights $(w, \hat w)$, transforms under $SL(2,\ZZ)$ as
\begin{equation}
f^{(w,\hat w)} (\gamma \cdot \tau,\gamma \cdot \bar{\tau}) = (c\tau+d)^w  \, (c\bar \tau +d)^{\hat w}\,  f^{(w, \hat w)} (\tau,\bar{\tau}) \,.
\label{modtrans}
\end{equation}
Modular covariant derivatives are defined by
\begin{equation}
 \cD_w= i\left(\tau_2 \frac{\partial}{\partial \tau} - i\,\frac{w}{2} \right),
\qquad  \bar{\cD}_{\hat w} = -i\left(\tau_2
\frac{\partial}{\partial \bar\tau} +i \frac{\hat w}{2}\right)\, , 
\label{covderdef}
\end{equation}
where $\cD_w$ transforms a modular form with weights $(w, \hat w)$ to a new modular form with $(w+1,\hat w-1)$ and $\bar \cD_{\hat w}$ changes weights by $(w, \hat w)\to (w-1, \hat w+1)$, i.e. $\cD_w: M_{w,\hat{w}} \mapsto M_{w+1,\hat{w}-1}$ and similarly $\bar \cD_{\hat w}: M_{w,\hat{w}} \mapsto M_{w-1,\hat{w}+1}$.  In other words,
\begin{equation}
\label{faction}
\cD_w\,  f^{(w, \hat w)} (\tau,\bar{\tau}):= f^{(w+1,\hat w-1)}(\tau,\bar{\tau})\,,\qquad\quad \bar \cD_{\hat w}\, f^{(w, \hat w)}(\tau,\bar{\tau}) := f^{(w-1, \hat w+1)}(\tau,\bar{\tau}) \, .
\end{equation}
Non-holomorphic forms for which $\hat w=-w$, are particularly  relevant to our discussion and transform by a phase characterised by a $U(1)$ charge, $q=2w$, as is evident from  (\ref{modtrans}). It is useful   to note that the action of $\cD_w$ on a power of $\mbox{Im}\tau = \tau_2=1/g_s$ is given by 
\begin{equation}
  \cD_w \tau_2^\alpha =  \frac{1}{2}\left( \tau_2 \frac{\partial}{\partial \tau_2} +w\right)\tau_2^\alpha
=  \frac{1}{2} \left(-g_s \frac{\partial}{\partial g_s} +w\right)  g_s^{-\alpha}\,.
\label{tauind}
 \end{equation}

The operators $ \cD_w, \bar \cD_{\hat w}$, together with the Cartan operator $H_{w,\hat{w}} =( w-\hat{w})/2:M_{w,\hat{w}}\mapsto M_{w,\hat{w}}$, form a representation of the $\mathfrak{sl}(2)$ algebra on $M_{w,\hat{w}}$. 
The Casimir operator for this representation yields Laplace-like differential operators which map $M_{w,\hat{w}}$ into itself. In particular, restricting to the case $\hat{w}=-w$, we have the Laplacians  
\begin{align}\label{eq:Lap}
\Delta_{(-)w}:=4  \cD_{w-1} \bar \cD_{-w}\,,\\
\Delta_{(+)w}:=4   \bar \cD_{-w-1}\cD_{w}\,.
\end{align}
Note that in the modular invariant case $M_{0,0}$ these reduce to the standard Laplacian $\Delta_{(-)0} = \Delta_{(+)0} := \Delta_{\tau} = 4\tau_2^2 \partial_\tau\partial_{\bar{\tau}}  $.
 
Homogeneous (and inhomogenous) Laplace eigenvalue equations on the space $M_{w,-w}$  arise at various stages in this review.  These have the equivalent forms 
\bea
\label{laplaceone}
& \Delta_{(-) w} \, f_s^{(w,-w)}(\tau,\bar{\tau}) = \left(s(s-1)-w(w-1)\right)\, f_s^{(w,-w)}(\tau,\bar{\tau})\, , \cr
& \Delta_{(+) w} \, f_s^{(w,-w)}(\tau,\bar{\tau}) = \left(s(s-1)-w(w+1)\right)\, f_s^{(w,-w)}(\tau,\bar{\tau})\, ,
\eea
where $s\in \mathbb{C}$.
These equations have a unique solution in $M_{w,\hat{w}}$, for functions satisfying the physically required boundary condition of moderate growth (power behaviour) in the large-$\tau_2$ limit (the weak-coupling limit).  

A basic ingredient in our discussion is the modular invariant non-holomorphic Eisenstein series, which is defined by
\bea
\label{eq:eisendef}
E(s; \tau, \bar{\tau}) =   \frac{1}{\pi^s} \sum_{(m,n) \neq (0,0)} {\tau_2^s \over |m+n \tau|^{2s}} \, ,\qquad E(s; \tau, \bar{\tau}) \in M_{0,0}\,,
\eea
and satisfies the homogeneous Laplace eigenvalue equation
\bea 
 \label{eq:Es}
\left[ \Delta_{\tau} - s(s-1) \right] E(s; \tau, \bar{\tau})  =0\, .
\eea
It has a well-known Fourier mode decomposition, 
\bea \label{eq:Est}
E(s; \tau, \bar{\tau}) = \sum_{k \in \ZZ} \mathcal{F}_k(s; \tau_2) e^{2\pi i k \tau_1} \, ,
\eea
 where zero mode (or equivalently the perturbative term) is given by
\bea \label{eq:Es-per}
\mathcal{F}_0(s; \tau_2) &= {2\zeta(2s) \over \pi^s} \tau_2^s + {2\sqrt{\pi} \Gamma(s-\half) \zeta(2s-1) \over \pi^s \Gamma(s)} \tau_2^{1-s} \, , 
\eea
and the non-zero modes are given by 
\bea
\mathcal{F}_k(s; \tau_2) &= {4\over \Gamma(s)} |k|^{s-\half} \sigma_{1-2s}(|k|) \sqrt{\tau_2} K_{s-\half}(2\pi |k| \tau_2) \, , \quad k\neq 0 \, ,  
\eea
where the divisor sum is defined by $\sigma_\nu(k) = \sum_{d|k} d^{\nu}$.  The non-zero mode  $\mathcal{F}_k$ represents the $k$-instanton contribution. 

Our discussion also involves non-holomorphic $(w,-w)$-forms, $E_w(s; \tau, \bar{\tau})\in M_{w,-w}$,  that are defined by 
\bea 
\label{eq:lattice}
E_w(s; \tau, \bar{\tau}) &= & {2^w \Gamma(s) \over \Gamma(s+w)} \cD_{w-1} \cdots \cD_0 E(s; \tau, \bar{\tau})  
\nonumber \,,\\
& =& {1\over \pi^s} \sum_{(m,n) \neq (0,0)} \left( {m+n \bar{\tau}\over m+n {\tau} } \right)^w {\tau_2^s \over |m+n \tau|^{2s}} \, ,
\eea
(where $E_0(s; \tau, \bar{\tau}) =E(s; \tau, \bar{\tau}) $)  which satisfy the recursion relations, 
\begin{align}
\cD_w E_w(s; \tau, \bar{\tau}) &= {s+w\over 2} E_{w+1}(s; \tau, \bar{\tau}) \, ,\\
  \bar\cD_{-w} E_w(s; \tau, \bar{\tau}) &= {s-w\over 2} E_{w-1}(s; \tau, \bar{\tau})\, .
\end{align}

Another type of modular function that plays an important r\^ole in this article is the generalised non-holomorphic Eisenstein series, which satisfies the inhomogenous Laplace eigenvalue equation, 
\bea 
\label{eq:generalise}
\left[ \Delta_{\tau} - r(r+1) \right] \mathcal{E}(s_1, s_2; r;  \tau,\bar{\tau})  =  -E(s_1; \tau,\bar{\tau})E(s_2; \tau,\bar{\tau}) \, .
\eea
 This equation again has a unique  $SL(2, \ZZ)$-invariant solution given appropriate boundary conditions. The prototype of this equation arises in considering the coefficient of  $d^6R^4$ in the effective type IIB action (\ref{eq:LLeff})  where $s_1=s_2=3/2$ and $r=3$\cite{Green:2005ba,Green:2014yxa}.
The complete solutions of this equation for generic $s_1, s_2, r\in\mathbb{C}$ are not known,  although a spectral decomposition has been studied \cite{Klinger-Logan:2018sjt}. 

However, more is known about the solutions to (\ref{eq:generalise}) when $s_1,s_2$ are integers or when they are half-integers.  Both of these cases are of special relevance to this article.
The complete perturbative and non-perturbative expansions  were obtained in \cite{Dorigoni:2021jfr,Dorigoni:2021ngn}  when  $s_1,s_2\in \mathbb{N}$, which plays a r\^ole in the study of the low-energy expansion of genus-one Type II superstring amplitudes, here discussed in section \ref{sec:otherasp}.
Similarly, cases with $s_1,s_2 \in \mathbb{N}+ 1/2$ (which  includes  the special case, $\mathcal{E}(\threeh, \threeh; 3; \tau)$)  were discussed in  \cite{Green:2014yxa,Green:2008bf}.    

For illustrative purposes, as well as because of its relevance to (\ref{eq:LLeff}), we will sketch the form of $\mathcal{E}(\threeh, \threeh; 3; \tau,\bar{\tau} )$, which differs significantly from that of a non-holomorphic Eisenstein series.  Its zero Fourier mode  is given by
\bea 
\label{eq:ME-per}
\!\!\!\!\!\!\!\!\!\!\!\!\!\!\!\!\!\!\!\!\!\!\!\!\! \mathcal{E}(\threeh, \threeh; 3; \tau,\bar{\tau}) |_{zero\ mode} &&\nonumber \\
&&   
\!\!\!\!\!\!\!\!\!\!\!\!\!\!\!\!\!\!\!\!\!\!\!\!\!\!\!\!\!\!\!\!\!\!\!\!\!\!\!\!\!\!\!\!\!\!\!\!\!\!= {2\over 3} \zeta(3)^2 \tau_2^3 + {4\over 3} \zeta(2)\zeta(3) \tau_2
+ {8\over 5} \zeta(2)^2 \tau_2^{-1} +{4\over 27} \zeta(6)  \tau_2^{-3} + O(e^{-4\pi \tau_2}) \,,
\eea
where $O(e^{-4\pi \tau_2})$  indicates the presence of an infinite series of powers of $(q \bar q)$, where $q=e^{2\pi i \tau}$  (and recalling that $\tau_2=1/g_s$, where $g_s$ is the string coupling constant).  These terms are interpreted as contributions of  instanton anti-instanton pairs.
The power-behaved terms are interpreted as coefficients of perturbative contributions in the $d^6R^4$ term in the low-energy expansion of four-point amplitude in  the superstring theory. 

Finally, we note that, starting from $\mathcal{E}(s_1,s_2; r; \tau,\bar{\tau})$ one can construct weight-$(w,-w)$  modular forms, which also play a r\^ole in this article,  by acting with covariant derivatives 
\bea \label{MEw}
\mathcal{E}_w(s_1, s_2; r; \tau,\bar{\tau}) \equiv {1\over 2^{w}} \cD_{w-1} \cdots \cD_1 \cD_0 \, \mathcal{E}(s_1, s_2; r; \tau,\bar{\tau})\, ,
\eea
similar to \eqref{eq:lattice}.

\vspace{0.2cm}
\begin{center}{\rule{8cm}{0.3mm}}\end{center}
\vspace{0.4cm}

\providecommand{\newblock}{}

\end{document}